\pdfoutput=1
\documentclass[fleqn,usenatbib]{mnras}

\usepackage[T1]{fontenc}
\usepackage{ae,aecompl}

\usepackage{graphicx}	
\usepackage{amsmath}	
\usepackage{amssymb}	
\usepackage[usenames,dvipsnames]{xcolor}
\usepackage{hyperref}
\usepackage[normalem]{ulem}
\usepackage{soul}

\newcommand{\sub}[1]{_{\mathrm{#1}}}
\newcommand{\msun}{M\sub{\sun}}

\def\equationautorefname~#1\null{Eq.~(#1)\null}
\def\figureautorefname~#1\null{Fig.~#1\null}
\newcommand{\appref}[1]{\hyperref[#1]{Appendix~\ref{#1}}}
\usepackage{newtxtext,newtxmath}

\title[Subhalo orbits in smoothly-growing host haloes]{Evolution of subhalo orbits in a smoothly-growing host halo potential}

\author[G. Ogiya et al.]{
Go Ogiya$^{1,2}$\thanks{E-mail: gogiya@uwaterloo.ca (GO)}, 
James E. Taylor$^{1,2}$, and 
Michael J. Hudson$^{1,2}$
\\
$^{1}$Waterloo Centre for Astrophysics, University of Waterloo, Waterloo, ON N2L 3G1, Canada \\
$^{2}$Department of Physics and Astronomy, University of Waterloo, 200 University Avenue West, Waterloo, Ontario N2L 3G1, Canada
}

\date{Accepted XXX. Received YYY; in original form ZZZ}

\pubyear{2020}

\begin{document}
\label{firstpage}
\pagerange{\pageref{firstpage}--\pageref{lastpage}}
\maketitle

\begin{abstract}
The orbital parameters of dark matter (DM) subhaloes play an essential role in determining their mass-loss rates and overall spatial distribution within a host halo. Haloes in cosmological simulations grow by a combination of relatively smooth accretion and more violent mergers, and both processes will modify subhalo orbits. To isolate the impact of the smooth growth of the host halo from other relevant mechanisms, we study subhalo orbital evolution using numerical calculations in which subhaloes are modelled as massless particles orbiting in a time-varying spherical potential. We find that the radial action of the subhalo orbit decreases over the first few orbits, indicating that the response to the growth of the host halo is not adiabatic during this phase. The subhalo orbits can shrink by a factor of $\sim$1.5 in this phase. Subsequently, the radial action is well conserved and orbital contraction slows down. We propose a model accurately describing the orbital evolution. Given these results, we consider the spatial distribution of the population of subhaloes identified in high-resolution cosmological simulations. We find that it is consistent with this population having been accreted at $z \la 3$, indicating that any subhaloes accreted earlier are unresolved in the simulations. We also discuss tidal stripping as a formation scenario for NGC1052-DF2, an ultra diffuse galaxy significantly lacking DM, and find that its expected DM mass could be consistent with observational constraints if its progenitor was accreted early enough, $z \ga 1.5$, although it should still be a relatively rare object. 
\end{abstract}




\section{Introduction}
\label{sec:intro}

In the standard paradigm for structure formation in the Universe, the $\Lambda$ cold dark matter ($\Lambda$CDM) cosmological model, small dark matter (DM) haloes form early on through gravitational collapse, and then merge to form larger structures subsequently. At the same time, baryonic gas cools within haloes over some mass scale, igniting galaxy formation. As a result, the hierarchy of observed galaxies is formed \citep[e.g.,][]{White1978,Frenk2012}, i.e., smaller DM subhaloes and associated satellite galaxies, orbiting within larger host systems consisting of a host DM halo and a central galaxy. In this paper, we consider the orbital evolution of subhaloes in their host halo, given the mass assembly history of the latter. 

The orbital parameters of subhaloes are essential in determining their spatial distribution and mass-loss rates, and thus they are an important input to observational tests of the nature of DM. For instance, in a structure formation scenario based on an alternative DM model, warm dark matter (WDM), the primordial density fluctuations on small scales are smoothed out by the free-streaming motion of WDM particles, and fewer subhaloes are formed, while the minimum halo mass is greater than in the $\Lambda$CDM cosmology \citep[][and references therein]{Bode2001,Angulo2013,Lovell2014,Bose2016}. This should affect both the abundance and the spatial distribution of subhaloes, since the formation of the earliest, densest structures would be suppressed. Tests of DM  based on observations of gravitational lensing \citep[e.g.,][]{Dalal2002,Vegetti2012,Shu2015,Hezaveh2016}, gaps in stellar streams \citep[e.g.,][]{Carlberg2012,Ngan2014,Erkal2016,Ibata2020}, and annihilation or decay signals of DM particles \citep[e.g.,][]{Strigari2007,Pieri2008,Hayashi2016,Hiroshima2018, Okoli2018} also depend strongly on the expected abundance, spatial distribution and mass function of DM subhaloes.

Subhaloes evolve dynamically within the host halo. Since subhaloes are extended objects, the gravity of the host halo works as a tidal force and gradually strips a subhalo's mass \citep[e.g.,][]{King1962,Spitzer1987,Taylor2004,Binney2008}. The subhalo orbit plays an essential role in determining the dynamical evolution of subhaloes in the tidal force field. When a subhalo is on a radial orbit, its mass is significantly reduced by tidal stripping once per orbit, at the pericentre where it feels the strongest tidal force. On the other hand, a subhalo's mass is reduced more gradually when it is on a circular orbit. As a response to tidal stripping, the internal structure of the subhalo is altered through re-virialization \citep[e.g.,][]{Hayashi2003,Penarrubia2010,Drakos2017,Ogiya2019,Delos2019,Green2019,Drakos2020}. An accurate prediction of this tidal evolution is required to interpret observational tests of the nature of DM.

When the host halo is spherical and static, the orbit of a subhalo is specified by its energy $E$, and angular momentum vector, or equivalently by the energy, orbital plane, and scalar angular momentum $L$. In this paper, we use two dimensionless parameters to characterise the orbit. The first one, characterising the orbital energy, is 
\begin{eqnarray}
    x\sub{c} \equiv r\sub{c}(E)/r\sub{200}, 
        \label{eq:xc}
\end{eqnarray}
where $r\sub{c}(E)$ and $r\sub{200}$ are the radius of a circular orbit of the orbital energy, $E$, and the virial radius of the host halo (see \autoref{eq:vir_mass} below). The second parameter, the orbital circularity, characterises the angular momentum of subhalo's orbit,
\begin{eqnarray}
    \eta \equiv L/L\sub{c}(E),
        \label{eq:eta}
\end{eqnarray}
where $L$ and $L\sub{c}(E)$ are the angular momentum of the subhalo orbit and the angular momentum of a circular orbit with the same energy. Orbits of $\eta=0$ and 1 are purely radial and circular, respectively. A number of authors have studied the orbital properties of subhaloes in cosmological $N$-body simulations  \citep[e.g.,][]{Tormen1997,Zentner2005,Khochfar2006,Wetzel2011,Jiang2015,vandenBosch2018a} and have shown that the distributions of these parameters peak at $x\sub{c} \sim 1.2$ and $\eta \sim 0.6$, respectively. 

These studies generally measure the orbital properties of subhaloes at the time of accretion, but their orbits will evolve subsequently, due to several different processes, including dynamical friction and self-friction. Dynamical friction is the drag force exerted by the wake that forms behind the subhalo in the density field of the host halo \citep{Chandrasekhar1943}, due to the subhalo's own gravity \citep[e.g.,][]{Ogiya2016}. Self-friction is another drag force caused by tidally stripped material from the subhalo \citep{Fujii2006,Fellhauer2007,vandenBosch2018b}. Similar to dynamical friction, self-friction works more efficiently when the ratio of the subhalo mass to the host mass is larger \citep{Ogiya2019,Miller2020}. Note that self-friction operates in all simulations associated with tidal mass-loss, even if the host halo is modelled with an analytical potential\footnote{In the simulations using of an analytical potential for the host, dynamical friction is absent because of the absence of the density wakes.}. These mechanisms couple with each other non-linearly and drive orbital evolution.

Due to the difficulty of developing fully analytical treatments for the underlying physics, semi-analytical modelling, in which fudge parameters in analytical formulae are calibrated to reproduce the simulation results, is a useful way to study the orbital evolution of subhaloes \citep[e.g.,][]{Lacey1993,Taylor2001,Taffoni2003,Penarrubia2005,Zentner2005,Boylan-Kolchin2008,Jiang2008,Gan2010,Pullen2014}. While the fudge parameters are mainly calibrated for the formulation of dynamical friction \citep{Chandrasekhar1943}, self-friction is also taken into account since the simulations are associated with tidal mass-loss of subhaloes. Using idealised simulations, \cite{Miller2020} found that the impact of self-friction is sub-dominant ($\sim 1/10$ of dynamical friction) in mergers between a host halo and a subhalo.

Other mechanisms, including violent relaxation driven by major mergers \citep{Lynden-Bell1967}, interactions between subhaloes \citep[e.g.,][]{Moore1996} and the growth of the host halo, make the potential field time-varying and can alter the orbits of subhaloes. While all of these, together with the drag forces of dynamical friction and self-friction, contribute to the orbital evolution of subhaloes, their individual impacts have been less well studied and are thus uncertain. The fact that these mechanisms are also coupled makes understanding the orbital evolution of subhaloes in cosmologically realistic situations particularly challenging.

In the hierarchical structure formation scenario, DM haloes grow through mergers with other haloes and smooth accretion from adjoining filaments or the surrounding field. Both processes change the potential of DM haloes with time, and alter the orbits of their subhaloes. To avoid the complexities raised by mergers, which introduce several new parameters and internal degrees of freedom to the problem, in this paper we focus on the impact of the smooth growth on subhalo orbits. We use numerical calculations in which a spherical host halo is modelled with an analytical potential, and the orbits of subhaloes are traced with massless particles. This treatment allows us to isolate the impacts of the smooth host halo growth from those of the other mechanisms.
 
The rest of the paper is organised as follows: \autoref{sec:sim_model} describes the setup and assumptions of our numerical models. In \autoref{sec:subhalo_orb_evo}, we study the orbital evolution of subhaloes driven by the smooth mass growth of the host halo, and derive an empirical model describing the evolution of subhalo orbits. We discuss the spatial distribution of subhaloes in \autoref{sec:spat_dist} and the mass evolution of the possible progenitor of the DM-deficient galaxy NGC1052-DF2 in \autoref{sec:df2}, before summarising our results in \autoref{sec:summary}. Throughout this paper, we use the cosmological parameters obtained by \cite{Planck2016}.

\section{Numerical Model}
\label{sec:sim_model}

We use numerical calculations to study the orbital evolution of subhaloes driven by the smooth growth of the spherical host halo potential. In these calculations, subhaloes are treated as massless particles, i.e.~interactions between subhaloes, dynamical friction and self-friction are all neglected, such that the (time-varying) analytic potential representing the host halo fully governs their motion after accretion. Therefore we solve $N$ individual one-body problems. The following part of this section describes the justification for, and details of, the numerical model.

Here, we define basic quantities in our model. The virial mass of the DM halo is given as 
\begin{eqnarray}
    M\sub{vir} \equiv \frac{4 \pi}{3} \Delta\sub{vir}(z) \rho\sub{crit}(z) r\sub{vir}^3, 
        \label{eq:vir_mass}
\end{eqnarray}
where $\Delta\sub{vir}(z)$ and $\rho\sub{crit}(z)$ are the virial overdensity and critical density of the universe at given redshift $z$. As seen in \autoref{eq:vir_mass}, the virial radius, $r\sub{vir}$, is the radius in which the mean density is $\Delta\sub{vir}(z)$ times of the critical density at $z$. The virial overdensity of $\Delta\sub{vir}(z)=200$ is employed, and hereafter we denote the virial mass and radius of the host halo as $M\sub{200}$ and $r\sub{200}$. Throughout this paper, the DM host halo is taken to be spherical, with a Navarro-Frenk-White (NFW) density profile \citep{Navarro1997},
\begin{eqnarray}
    \rho(r) = \frac{\rho\sub{s}}{(r/r\sub{s})(1+r/r\sub{s})^2}\,. 
        \label{eq:nfw}
\end{eqnarray}
Here $r$, $\rho\sub{s}$ and $r\sub{s}$ are the distance from the host halo's centre, the scale density,  and the scale length of the host halo, respectively. The halo concentration, $c$, is defined as 
\begin{eqnarray}
    c \equiv r\sub{200}/r\sub{s}. 
        \label{eq:concentration}
\end{eqnarray}

\subsection{Modelling subhalo orbits with massless particles}
\label{ssec:mass_less_particle}

A key assumption in our numerical calculations is treatment of subhaloes as massless particles, which corresponds to neglecting dynamical friction, self-friction and subhalo-subhalo interactions. Here we justify this approximation.

\subsubsection{Dynamical friction}
\label{sssec:dyn_fric}
The strength of dynamical friction and self-friction relative to gravity is roughly proportional to the mass of the subhalo \citep[e.g.,][]{Chandrasekhar1943,Miller2020}. For the low mass systems that dominate the subhalo population \citep[][and references therein]{Giocoli2008,Springel2008,Jiang2016}, these effects will be negligible, and thus treating subhalos as massless particles is justified. On the other hand, the orbits of larger subhaloes will shrink more than our numerical calculations predict, due to the neglected impacts of dynamical friction and self-friction.

What is the condition on subhalo mass, to be able to neglect these drag forces? Because self-friction only works subdominantly \citep{Miller2020}, we will only consider dynamical friction in what follows. According to Chandrasekhar's formula for dynamical friction, the deceleration is given as
\begin{eqnarray}
    {\bf a}\sub{DF} = -4 \pi G^2 \ln{\Lambda} \frac{\rho\sub{h} M\sub{s}}{v^2} \frac{{\bf v}}{v},
        \label{eq:adf}
\end{eqnarray}
where $G$, $\ln{\Lambda}$ and $\rho\sub{h}$ are the gravitational constant, the Coulomb logarithm, and the mass density of the host halo. A subhalo with a mass of $M\sub{s}$ moves in the host halo potential with velocity $v$. We define the orbital decay timescale of subhaloes accreted at $z\sub{acc}$ as
\begin{eqnarray}
    \tau(z\sub{acc}) &=& v/|a\sub{DF}| \nonumber \\
                     &=& f \bigl [ 2400 \pi G (\ln{\Lambda})^2 \rho\sub{crit}(z\sub{acc}) \bigr ]^{-1/2},
                        \label{eq:tau}
\end{eqnarray}
where $f \equiv M\sub{h}/M\sub{s}$. Here the host halo mass is given as $M\sub{h} \equiv M\sub{200}(z\sub{acc}) = (800 \pi/3) \rho\sub{crit}(z\sub{acc}) r\sub{200}^3(z\sub{acc})$ (\autoref{eq:vir_mass}) and we assume that the subhalo is on a circular orbit at the virial radius of the host halo at $z\sub{acc}$, $r\sub{200}(z\sub{acc})$. 

\begin{figure}
    \begin{center}
        \includegraphics[width=0.45\textwidth]{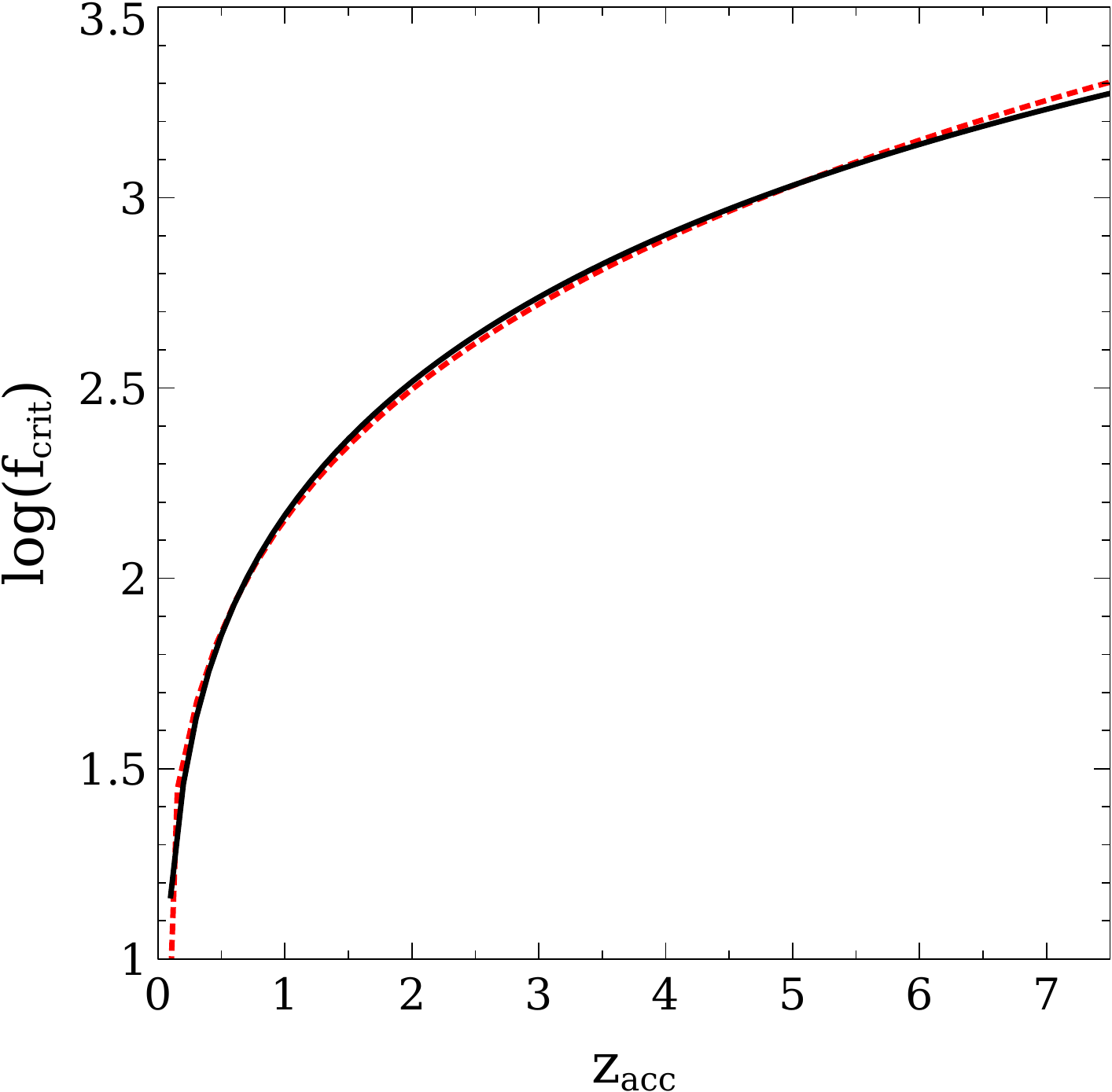}
    \end{center}
    \caption{
       The ratio of the host halo mass to the subhalo mass, $f\sub{crit}$, above which dynamical friction is negligible. The black curve is derived by equating the orbital decay time \autoref{eq:tau} to the lookback time for that accretion redshift, $z\sub{acc}$. A Coulomb logarithm of $\ln{\Lambda} = 5$ is assumed. The red dashed line represents the fitting function (\autoref{eq:fcrit_fit}). Since $f\sub{crit}$ decreases as $z\sub{acc}$ goes to zero, dynamical friction becomes less important at recent times. 
    \label{fig:fcrit}}
\end{figure}

Equating \autoref{eq:tau} to the lookback time corresponding to the accretion redshift, $z\sub{acc}$, we derive the critical ratio of the host halo mass to the subhalo mass, $f\sub{crit}(z\sub{acc})$. Dynamical friction and self-friction will be negligible for subhaloes of mass $M\sub{s} < M\sub{h}/f\sub{crit}$, while more massive systems will sink to the centre of the host halo and merge with it by the present time. In \autoref{fig:fcrit}, we show $f\sub{crit}$ as a function of $z\sub{acc}$ assuming $\ln{\Lambda}=5$ (black line). As shown by the red line, $f\sub{crit}$ is well fitted by 
\begin{eqnarray}
    \log{f\sub{crit}(z\sub{acc})} = 2.155 z\sub{acc}^{0.212} + \log{(\ln{\Lambda}/5)}.
                        \label{eq:fcrit_fit}
\end{eqnarray}
Because the lookback time is shorter and the critical density decreases at low redshift, $f\sub{crit}$ decreases as $z\sub{acc}$ goes to zero. Using \autoref{eq:fcrit_fit}, we find that dynamical friction and self-friction do not alter the orbit of subhaloes accreted at $z\sub{acc}=5$ (1) with a mass of $\sim 0.1$ (1) percent of the host halo mass at $z\sub{acc}$.

The impact of dynamical friction estimated in \autoref{eq:adf} is in fact an upper limit, given tidal mass-loss, which reduces the efficiency of dynamical friction. Thus the estimated timescale $\tau$ will be a lower limit, and $f\sub{crit}$ will be smaller than estimated. While the detailed mass-loss history for a subhalo will depend on its structure and orbit, $\tau$ gets three times longer in typical cases \citep{Mo2010} and thus $f\sub{crit}$ is reduced by a factor of three. In addition, only particles having velocity less than $v$ are actually expected to cause dynamical friction while \autoref{eq:adf} supposes all particles in the host halo contribute.

\subsubsection{Subhalo-subhalo interactions}
\label{sssec:subhalo_subhalo}
Interactions between subhaloes are another possible mechanism to drive orbital evolution, whose impact is neglected in our numerical calculations. To estimate how subhalo-subhalo interactions affect orbital evolution, we will use a toy model based on the argument in \S\,1.2.1 of \cite{Binney2008} The details of this model are described in \appref{app:model_subhalo_subhalo}.

\begin{figure}
    \begin{center}
        \includegraphics[width=0.45\textwidth]{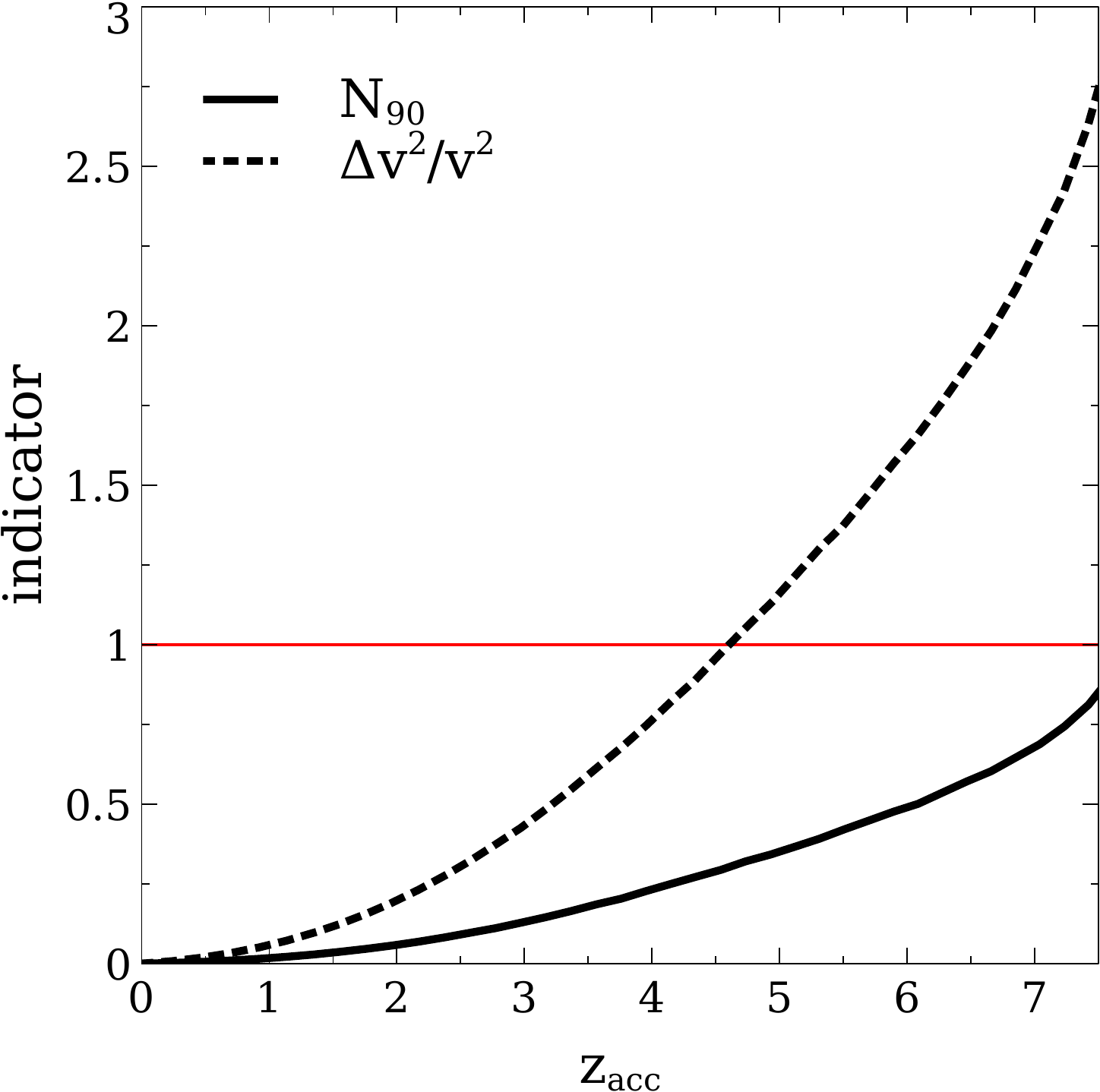}
    \end{center}
        \caption{
            The expected number of close encounters (solid) and the impact of cumulative weak encounters (dotted) as a function of the redshift at which the subhalo was accreted into the host halo. A final virial mass of $M\sub{0}=10^{12}\,\msun$ is assumed, but the results are insensitive to $M\sub{0}$. While strong deflections by subhalo-subhalo interactions are  not expected, any information about the orbits of subhaloes accreted at $z\sub{acc} \ga 5$ will be lost by the present time, due to the cumulative impacts of weak encounters. 
        \label{fig:subhalo_subhalo_interactions}}
\end{figure}
We consider two channels by which subhalo-subhalo interaction can alter the orbit of a given `subject' subhalo. The first one is the deflection of the subject subhalo's orbit by an angle of more than 90 degrees (a `strong' deflection) due to a  single close encounter with another subhalo (the `perturber'). The second channel is through the cumulative effect of many weak encounters. \autoref{fig:subhalo_subhalo_interactions} studies how many strong deflections are expected to occur, and how strong the effect of cumulative weak encounters should be, as a function of accretion redshift. We find that while strong deflections are generally unlikely (solid curve), the cumulative impact of many weak encounters may scatter the orbital parameters away from their initial values for subhaloes accreted at $z\sub{acc} \ga 5$

The impact of subhalo-subhalo interactions estimated here should represent an upper limit. More massive perturbers disturb the orbit of the subject subhalo more efficiently (see \appref{app:model_subhalo_subhalo}). The toy model assumes constant masses for the perturbers, while these are in fact reduced through tidal interactions with the host halo. On the other hand, the toy model neglects the effect of new perturbers accreted at $\la z\sub{acc}$. We expect that these perturbers would have little effect on a subject subhalo's orbit, however; given their larger orbits and longer orbital periods, the subject subhalo could respond adiabatically to any changes that they generate in the main potential. More detailed studies based on cosmological $N$-body simulations are needed to fully explore the effect of encounters on subhalo orbits, but overall we expect it to be minor in most cases.

\subsection{Growth of the host halo}
\label{ssec:host_growth}

\begin{figure}
    \begin{center}
        \includegraphics[width=0.45\textwidth]{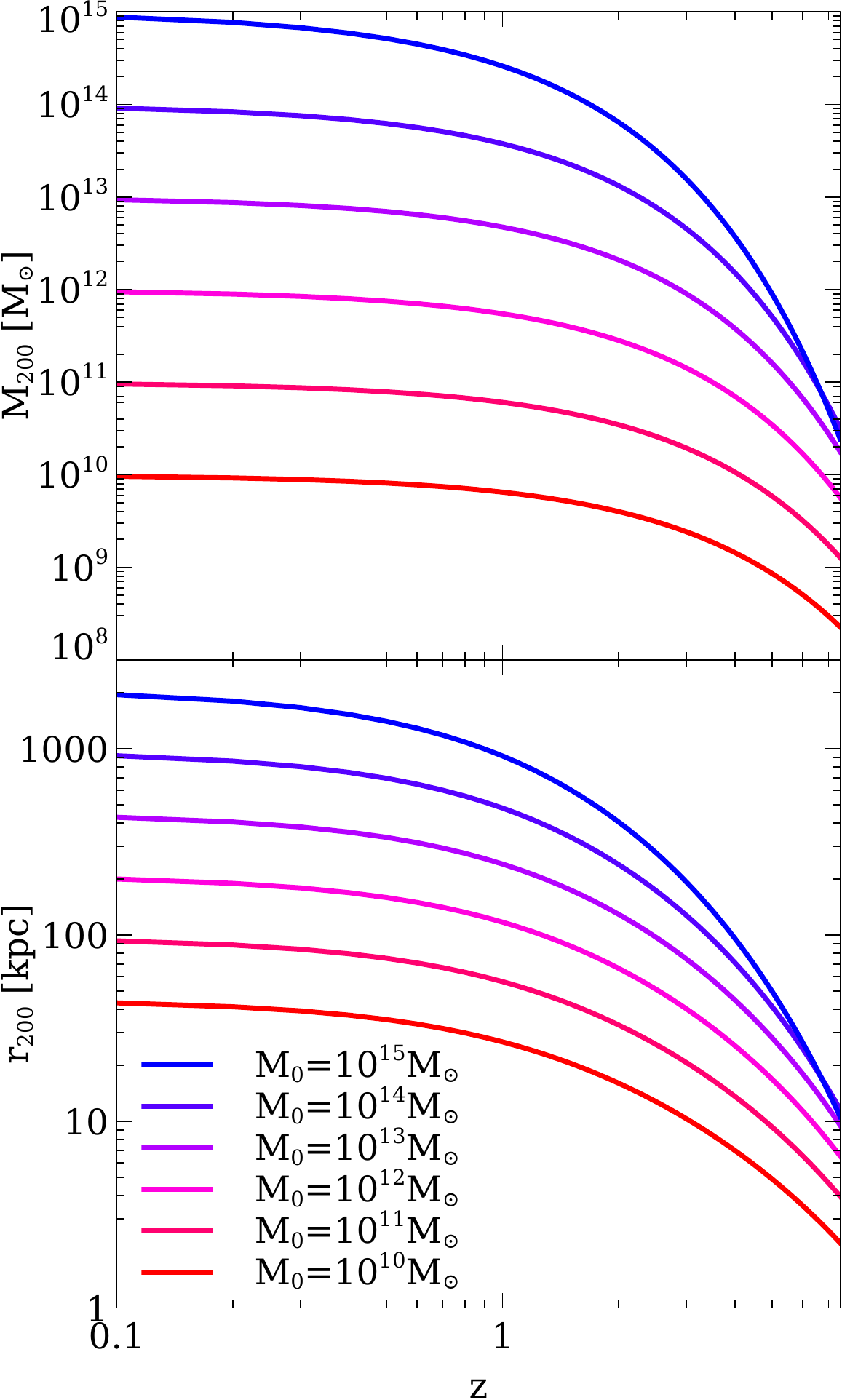}
    \end{center}
    \caption{
        The growth of the virial mass ({\it upper}) and radius ({\it lower}) of the host halo. Solid curves show models with a final virial mass of $M\sub{0}=10^{10},10^{11},10^{12},10^{13},10^{14}$ and $10^{15} \msun$. Lower (Higher) mass haloes grow earlier (later). 
    \label{fig:host_evo}}
\end{figure}

A number of previous studies have proposed models describing the mean mass assembly history (MAH) of DM haloes, based on the behaviour seen in cosmological $N$-body simulations \citep[e.g.,][]{Wechsler2002,McBride2009,Wong2012,vandenBosch2014}, or on extended Press-Schechter theory \citep[e.g.,][]{Lacey1993,vandenBosch2002,Correa2015a}. In this paper, given a virial mass for the host halo at the present time, $M\sub{200}(z=0) \equiv M\sub{0}$, we use the model of \cite{Correa2015b} to evaluate its prior evolution. The model requires a final halo concentration at $z=0$ and we use the concentration-mass-redshift relation, $c(M,z)$ of \cite{Ludlow2016} to set this. \autoref{fig:host_evo} shows the evolution of host halo mass and virial radius over time. DM haloes with lower final mass (redder lines) grow earlier than those with higher final mass (bluer lines), consistent with the basic pattern of hierarchical structure formation seen in cosmological $N$-body simulations \citep[e.g.,][]{Fakhouri2010}.  

\begin{figure}
    \begin{center}
        \includegraphics[width=0.45\textwidth]{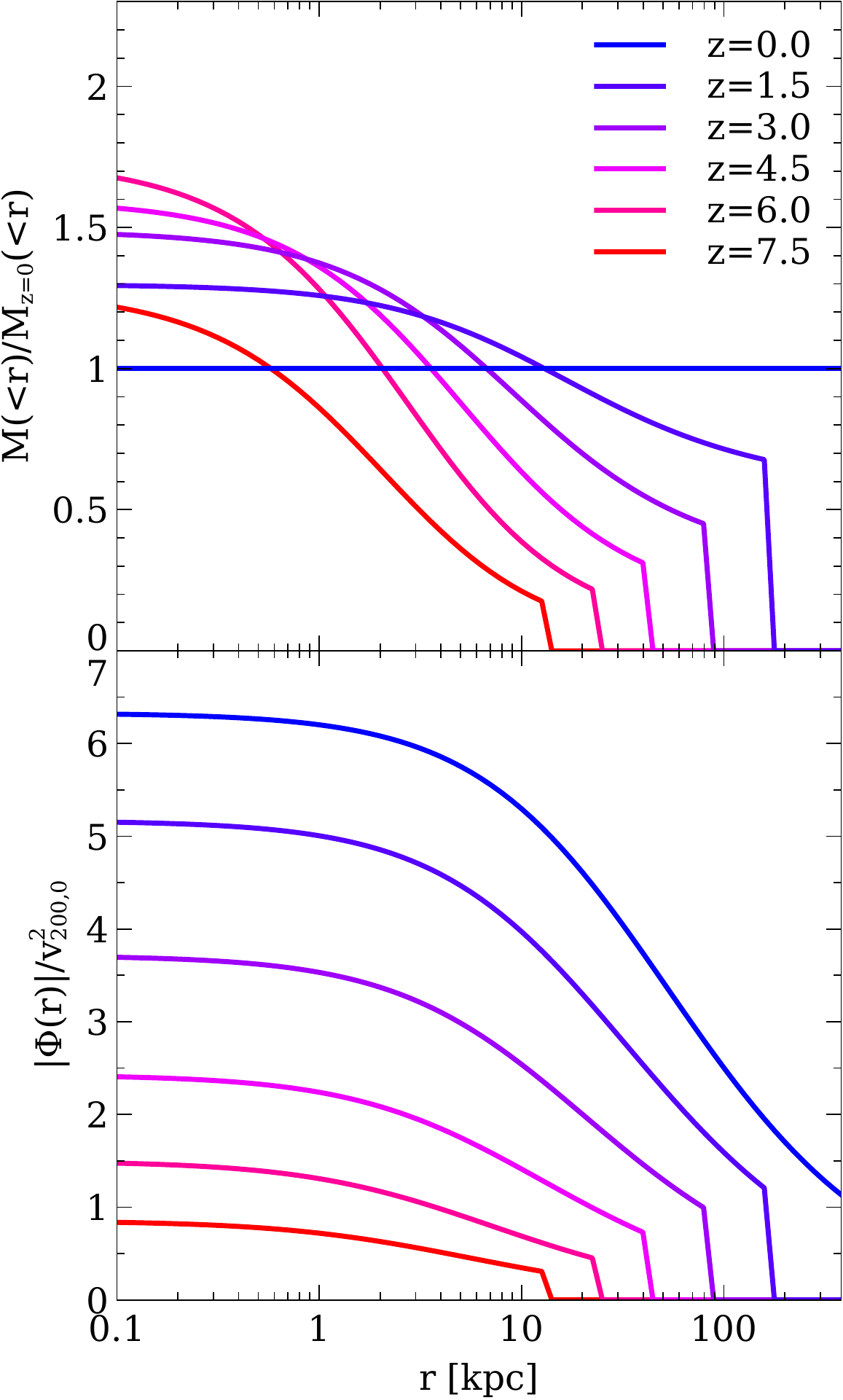}
    \end{center}
    \caption{
        Evolving mass and potential profiles of the model of $M\sub{0}=10^{12}\,\msun$. 
        ({\it Upper}) The enclosed mass profile at a given redshift $z$, $M(<r)$, relative to the profile at $z=0$, $M\sub{z=0}(<r)$. 
        ({\it Lower}) The potential profile at a given $z$, $\Phi(<r)$, relative to the virial velocity squared at $z=0$, $v\sub{200,0}^2 \equiv GM\sub{0}/r\sub{200}(z=0)$. 
        The radial bins are given in (fixed) physical kpc. The profiles at $z$ are computed in the range of $r=[0.1\,{\rm kpc}, 2\,r\sub{200}(z)]$ where $r\sub{200}(z)$ is the virial radius of the halo at $z$. 
        We note that the central density of the DM halo was higher at earlier times, peaking around $z\sim 6$, while the mass density at the outskirts increases at later times. The potential grows more overall at small radii than it does at the outskirts of the halo. 
    \label{fig:profile_fiducial}
    }
\end{figure}

Based on the model outlined above, we show the evolution of the radial profiles of the enclosed mass, $M(<r)$, and the gravitational potential, $\Phi(r)$, of the host halo with $M\sub{0}=10^{12}\,\msun$ in \autoref{fig:profile_fiducial}. Note that the radial bins are given in (fixed) physical kpc. In the upper panel, we see that at the outskirts of the halo, close to the virial radius (located roughly at the break point in the profiles), the enclosed mass $M(<r)$  increases steadily with time. On the other hand, close to the centre of the halo, the enclosed mass increases only until $z=6$, and then decreases with time. This behaviour seems inconsistent with some previous analyses of individual halos in cosmological $N$-body simulations, notably \citep[e.g.,][]{Diemand2007}. The DM halo analysed in \citet{Diemand2007} was isolated, however (with a last major merger at $z=1.7$), while DM haloes with more violent merger histories would be included in the derivation of the MAH model and $c(M,z)$ relation that our calculation assumes. The presence or absence of major mergers may explain the difference between the two pictures. We have tested several variants of the MAH model and/or $c(M,z)$ relation (\appref{app:profile_evolution}), but the evolution of the central mass structure is always qualitatively consistent with that shown in \autoref{fig:profile_fiducial}. The mass growth deepens the overall halo potential, with more growth in the centre than in the outskirts (lower panel). The potential of NFW haloes is relatively flat in the centre of the halo. Because of this, the potential at $r \la r\sub{s}$ deepens almost as much as that at $r=0$. Similar results hold for the models with other values of $M\sub{0}$. The results shown in the lower panel are robust even if the other MAH models and/or $c(M,z)$ relations are employed (\autoref{fig:profile_other_models}).

\subsection{Subhalo orbits at the time of accretion}
\label{ssec:subhalo_init_orbit}

In our numerical calculations, the host halo centre is fixed at the origin and the host-centric coordinates are used in what follows. Massless particles (subhaloes) fall into the host halo potential at the accretion redshift, $z\sub{acc}$. The host halo potential at $z\sub{acc}$ and the two orbital parameters, $x\sub{c}$ and $\eta$ (\autoref{eq:xc} and \autoref{eq:eta}) specify the orbital energy and angular momentum of massless particles at accretion. In the numerical calculations presented in \autoref{sec:subhalo_orb_evo}, we linearly sample them over the range of $x\sub{c}=[0.5:2]$ and $\eta=[0.05:0.95]$ in 10 steps of $\Delta x\sub{c}=0.15$ and $\Delta \eta = 0.09$ (i.e.~121 combinations at given $z\sub{acc}$). The two-dimensional parameter space covers the most of the initial subhalo orbits found in cosmological $N$-body simulations \citep[e.g.,][]{Tormen1997,Zentner2005,Khochfar2006,Wetzel2011,Jiang2015,vandenBosch2018a}. When not specified, at $z=z\sub{acc}$, massless particles are located at the apocentre, $r\sub{a}$, with zero radial velocity; we determine the amplitude of the tangential velocity using the pair of orbital parameters and the host halo potential at $z\sub{acc}$. The polar and azimuthal angles at accretion are randomly drawn. The amplitudes of the azimuth and polar components of the particle's velocity ($v\sub{\theta}$ and $v\sub{\phi}$) are randomly determined and satisfy $\eta L\sub{c}(E)/r\sub{i} = (v\sub{\theta}^2+v\sub{\phi}^2)^{1/2}$, where $r\sub{i}$ is the initial distance of the subhalo to the host halo centre.

\subsection{Numerical parameters}
\label{ssec:num_param}
We perform the numerical calculations from $z=7.5$ to 0, varying the virial mass of the host halo at $z=0$, $M\sub{0}$. $M\sub{0}$ is sampled over the range from $\log{[M\sub{0}/\msun]}=10$ to 15 in 15 steps of $\Delta \log{[M\sub{0}/\msun]}=1/3$ (i.e.~16 models in total). In each model, 121 massless subhalo particles accrete into the analytic host halo potential in redshift steps of $\Delta z=0.1$. Note that the number of massless particles in the numerical calculations does not reflect the actual number of subhaloes in a typical halo; instead, the numerical calculations sample subhalo orbits in the four dimensional space of i) accretion redshift, $z\sub{acc}$; i\hspace{-.1em}i) final mass of the host halo, $M\sub{0}$; i\hspace{-.1em}i\hspace{-.1em}i) orbital energy parameter at accretion, $x\sub{c,i}$; and i\hspace{-.1em}v) orbital circularity at accretion, $\eta\sub{i}$. In the end, we have 76 redshift bins and the total number of data points is $\sim 5.6 \times 10^6$. Neglecting interactions between subhaloes, computing the gravity of the spherical host halo potential is straightforward. We update the host halo mass and structure based on the models of \cite{Correa2015b} and \cite{Ludlow2016} every $\Delta t \approx 4.38 \times 10^5$\,yr. The particle orbits are integrated numerically using a second-order Leapfrog scheme with the same fixed timestep, $\Delta t$. Numerical calculations varying $\Delta t$, which control the smoothness of the host halo growth and the accuracy of orbit integration, confirm the numerical convergence of the results. We have also checked that the results are insensitive to the choice of $\Delta z$, $\Delta \log{[M\sub{0}/\msun]}$, $\Delta x\sub{c}$ and $\Delta \eta$.

\section{Orbital evolution of subhaloes in the smoothly-growing potential}
\label{sec:subhalo_orb_evo}

\subsection{How adiabatic is the host halo growth for subhalo orbits?}
\label{ssec:how_adiabatic}
If the mass growth of the spherical host halo is adiabatic, i.e.~the timescale for this growth is long compared to the typical orbital period for subhaloes, the evolution of the subhalo's orbit is specified analytically, based on the conserved actions, which are adiabatic invariants. Models based on the adiabatic invariants provide a reasonable description of the properties of observed galaxies \citep[e.g.,][]{Blumenthal1986,Dutton2007}, as well as those of galaxies in cosmological hydrodynamical simulations \citep[e.g.,][]{Gnedin2004}. One of the adiabatic invariants is the angular momentum of the subhalo orbit $L$; this is perfectly conserved in our numerical calculations because of the assumption of a spherical host halo, and the use of massless particles. The other one is the radial action, 
\begin{eqnarray}
    J\sub{r} = \oint\sub{orbit} v\sub{r}(r') dr' 
        \label{eq:jr}
\end{eqnarray}
where $v\sub{r}(r)$ is the radial velocity of the subhalo at $r$.

While $L$ and $J\sub{r}$ are the actual adiabatic invariants for spherical systems, their proxies are usually used to discuss the adiabatic evolution of DM and stellar orbits. The standard model by \cite{Blumenthal1986} assumes that a test particle moves on a circular orbit in an evolving spherical system; the proxy, $[GM(<r)r]^{1/2}$, corresponds to the specific angular momentum of the test particle, which is conserved in the evolution. However, the assumption of circular orbits is clearly unrealistic for most cases. \cite{Gnedin2004} proposed instead a modified proxy for $J\sub{r}$, $K \equiv [G M(<{\bar r}) {\bar r}]^{1/2}$, where ${\bar r}$ is the orbit-averaged radius,
\begin{eqnarray}
    {\bar r} = \oint\sub{orbit} \frac{r'dr'}{v\sub{r}(r')} \biggl / \oint\sub{orbit} \frac{dr'}{v\sub{r}(r')}
        \label{eq:rbar}.
\end{eqnarray}
While they found that the modified form, $K' = [G M(<{\bar r}) r]^{1/2}$, was a better proxy in their simulations, in our calculations we find that $K'$ can fluctuate by a factor of $\ga 10$ over a single orbital period, because of the variation in the instantaneous value of $r$. Thus, in the discussion below, $K$ is used. 

We will also consider another proxy of the adiabatic invariants, $L\sub{c} = [G M(<r\sub{c})r\sub{c}]^{1/2}$. The advantage of using $L\sub{c}$ is that it does not require orbit integration, unlike $K$, and we can evaluate it at any arbitrary phase of the orbit. Since $r\sub{c}$ roughly corresponds to the apocentre, $r\sub{a}$, $L\sub{c}$ is an approximation of another proxy considered by \cite{Blumenthal1986}, $GM(<r\sub{a})r\sub{a}$. This corresponds to $J\sub{r}^2$ when the orbit is purely radial. 

A way to estimate the adiabaticity of the orbital evolution of subhaloes is to compare the timescale of the change in the host halo potential, $t\sub{\Phi}$, to the orbital period for subhaloes, $t\sub{orb}$. The former is given as
\begin{eqnarray}
    t\sub{\Phi}(z,r) \equiv \Phi(z,r)/{\dot \Phi(z,r)},
        \label{eq:tphi}
\end{eqnarray}
where $\Phi(z,r)$ is the potential profile of the host halo at redshift, $z$, and ${\dot \Phi(z,r)}$ is its time derivative. The latter is proportional to the dynamical time of the halo, 
\begin{eqnarray}
    t\sub{dyn}(z,r) = \sqrt{\frac{3 \pi}{16 G {\bar \rho}(z,r)}} = \frac{\pi}{2} \sqrt{\frac{r^3}{GM(z,r)}},
        \label{eq:tdyn}
\end{eqnarray}
where ${\bar \rho}(z,r) = 3M(z,r)/(4 \pi r^3)$ is the mean density within $r$. We parametrise the ratio of $t\sub{orb}$ to $t\sub{dyn}$ as $\alpha$. According to \cite{Ogiya2019}, $t\sub{orb} \sim 6.7x\sub{c}^{1.15}$\,Gyr at $z=0$. Putting a typical value at accretion, $x\sub{c}=1.2$, while comparing $t\sub{orb}$ with $t\sub{dyn}[0,r\sub{200}(0)]$, we derive $\alpha=3.64$. 

\begin{figure}
    \begin{center}
        \includegraphics[width=0.45\textwidth]{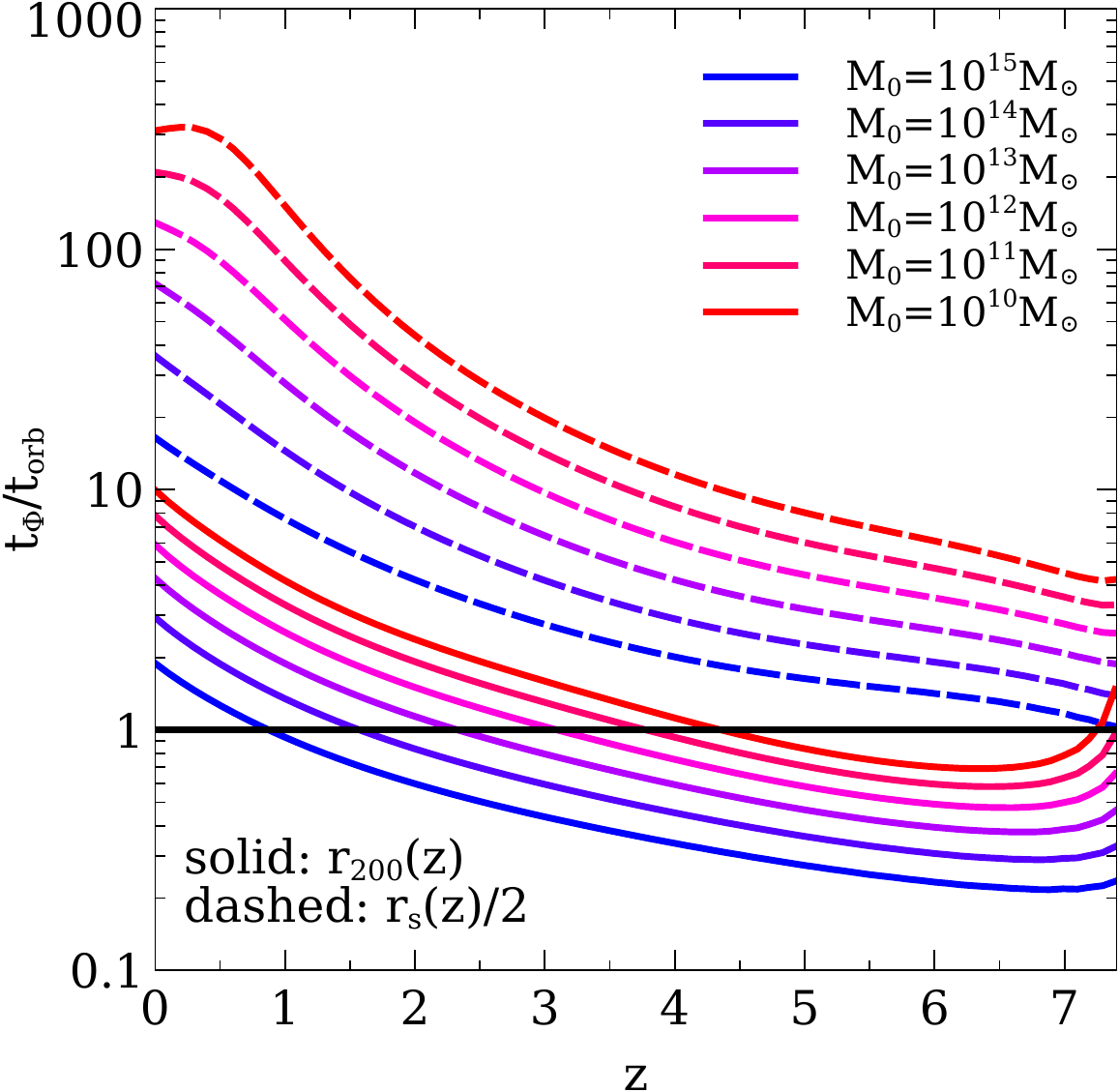}
    \end{center}
    \caption{
            Comparison between the timescale of the change in the potential, $t\sub{\Phi}$, and the orbital period, $t\sub{orb}$. Lines show the models with a final virial mass of $M\sub{0}=10^{10},10^{11},10^{12},10^{13},10^{14}$ and $10^{15} \msun$. The black horizontal line indicates equality between the two timescales, for guidance; above this line, the response to changes in the host potential should be adiabatic, while below the line it will  not be. For subhaloes orbiting close to the virial radius of the halo at redshift $z$, ($r\sub{200}(z)$; solid), $t\sub{\Phi}$ is shorter than or comparable to $t\sub{orb}$. Thus, the potential change is not adiabatic for recently accreted subhaloes with apocentres $\sim r\sub{200}(z)$. On the other hand, for subhaloes orbiting in the centre of the host halo ($r\sub{s}(z)/2$; dashed lines) the response to changes in the potential should be adiabatic at all redshifts.
    \label{fig:pot_change_time}}
\end{figure}

In \autoref{fig:pot_change_time}, we compare $t\sub{\Phi}$ to $t\sub{orb}$. Each line colour represents the results for a model with given value of $M\sub{0}$. Measuring the timescales at the halo outskirt ($r\sub{200}(z)$; solid), we find that $t\sub{\Phi} < t\sub{orb}$ at $z \ga 3$. While the ratio increases with decreasing of $z$, it is still of order unity in most cases. This indicates that the change in the host halo potential is not adiabatic for subhaloes orbiting around $r\sub{200}$, i.e.~ones recently accreted into the host halo. When the timescales are measured at the scale radius ($r\sub{s}(z)/2$; dashed), we find $t\sub{\Phi} \gg t\sub{orb}$ at all redshifts $z < 7$. Thus the growth of the host halo should produce adiabatic orbital changes for subhaloes in the halo centre. Such subhalos would generally have been accreted earlier, however, and would have experienced some non-adiabatic orbital evolution soon after accretion. 

We test the predicted adiabaticity of subhalo orbits with numerical calculations. To compute $J\sub{r}$ ($K$) in the $N\sub{p}$-th orbit, the numerical integration of \autoref{eq:jr} (\autoref{eq:rbar}) starts when the massless particle reaches the $N\sub{p}$-th apocentre, and continues until the massless particle reaches the $(N\sub{p}+1)$-th apocentre.  When evaluating the proxies $K$ and $L\sub{c}$, the mass profile of the host halo potential at the time of the $(N\sub{p}+1)$-th apocentre approach is used. 

\begin{figure}
    \begin{center}
        \includegraphics[width=0.45\textwidth]{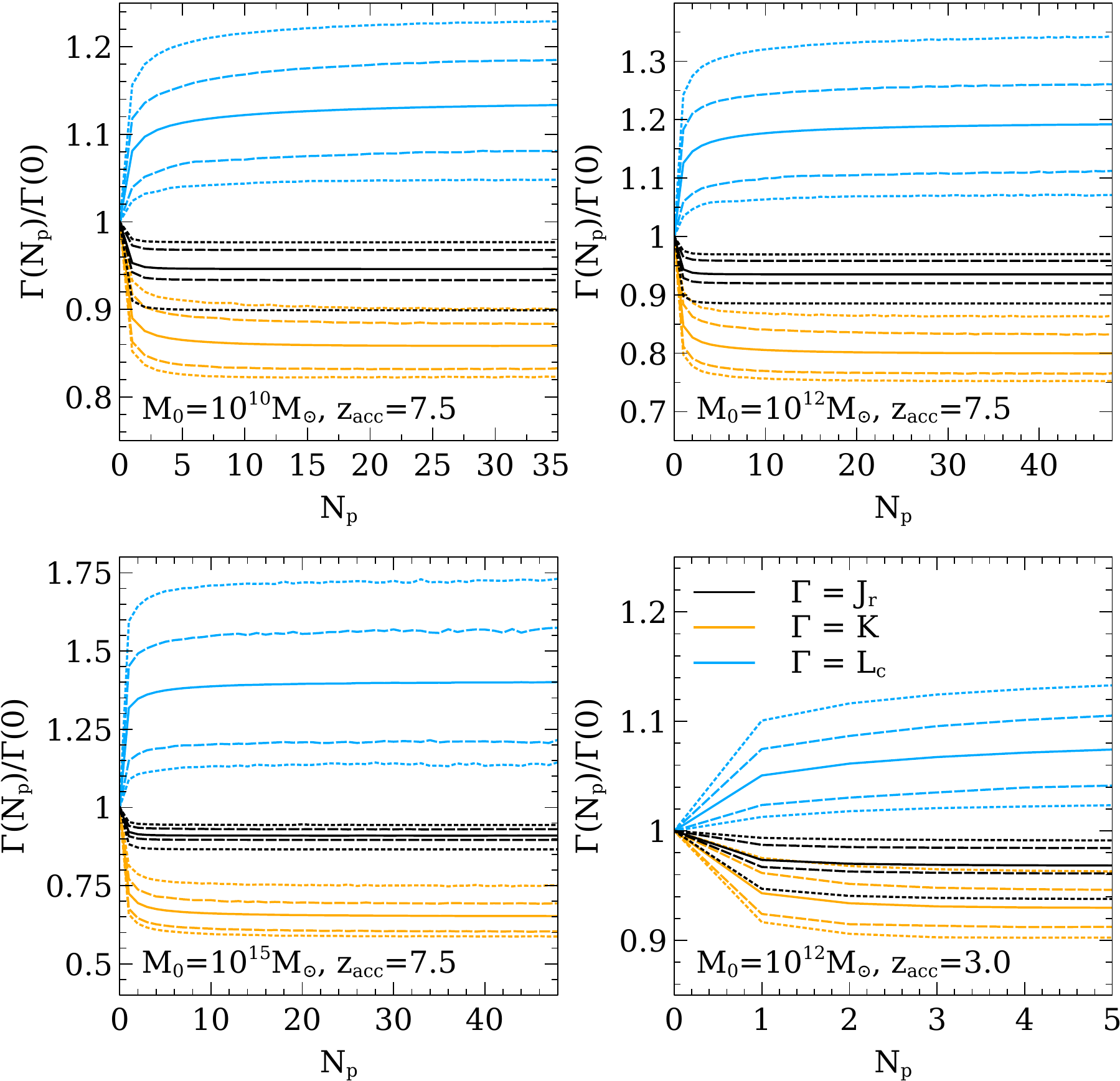}
    \end{center}
    \caption{
        Evolution of $J\sub{r}$ (black), $K$ (orange), and $L\sub{c}$ (blue), as a function of the number of orbital periods, $N\sub{p}$. Solid lines show the mean of the change in each quantity. Upper and lower dashed (dotted) lines show the 75th and 25th (90th and 10th) percentiles of the distribution. The host halo mass at $z=0$, $M\sub{0}$, and the accretion redshift, $z\sub{acc}$, are indicated in each panel. $J\sub{r}$ and its proxies change significantly over the first few orbits, but are well conserved subsequently.
    \label{fig:jr_and_proxies}}
\end{figure}
\autoref{fig:jr_and_proxies} presents the evolution of the adiabatic invariant $J\sub{r}$ (black), and its proxies $K$ (orange) and $L\sub{c}$ (blue), as a function of the number of orbital periods, $N\sub{p}$. We find that $J\sub{r}$ decreases by up to $\sim$\,10 percent over the first few orbital periods. This confirms that during this initial phase of evolution, the mass growth of the host halo potential is not adiabatic for subhalo orbits, so numerical calculations are needed to accurately model the orbital evolution of subhaloes in the growing host halo potential. In the later phase, $J\sub{r}$ remains constant, i.e.~the host halo growth appears adiabatic for subhalo orbits. These results are consistent with the expectation from the comparison of timescales in \autoref{fig:pot_change_time}. We also find in \autoref{fig:jr_and_proxies} that the change in the proxies in the non-adiabatic phase is greater than that in $J\sub{r}$, while they remain almost constant during the later adiabatic phase. Thus, the orbital evolution of subhaloes will be miscalculated if one relies on either of $K$ or $L\sub{c}$ evaluated at accretion. Also note that the net changes in $J\sub{r}$, $K$ and $L\sub{c}$ are larger for larger halo masses $M\sub{0}$ or earlier accretion times $z\sub{acc}$. In such cases, the ratio $t\sub{\Phi}/t\sub{orb}$ is smaller (\autoref{fig:pot_change_time}), i.e.~the change in the host halo potential is less adiabatic.

\begin{figure}
    \begin{center}
        \includegraphics[width=0.45\textwidth]{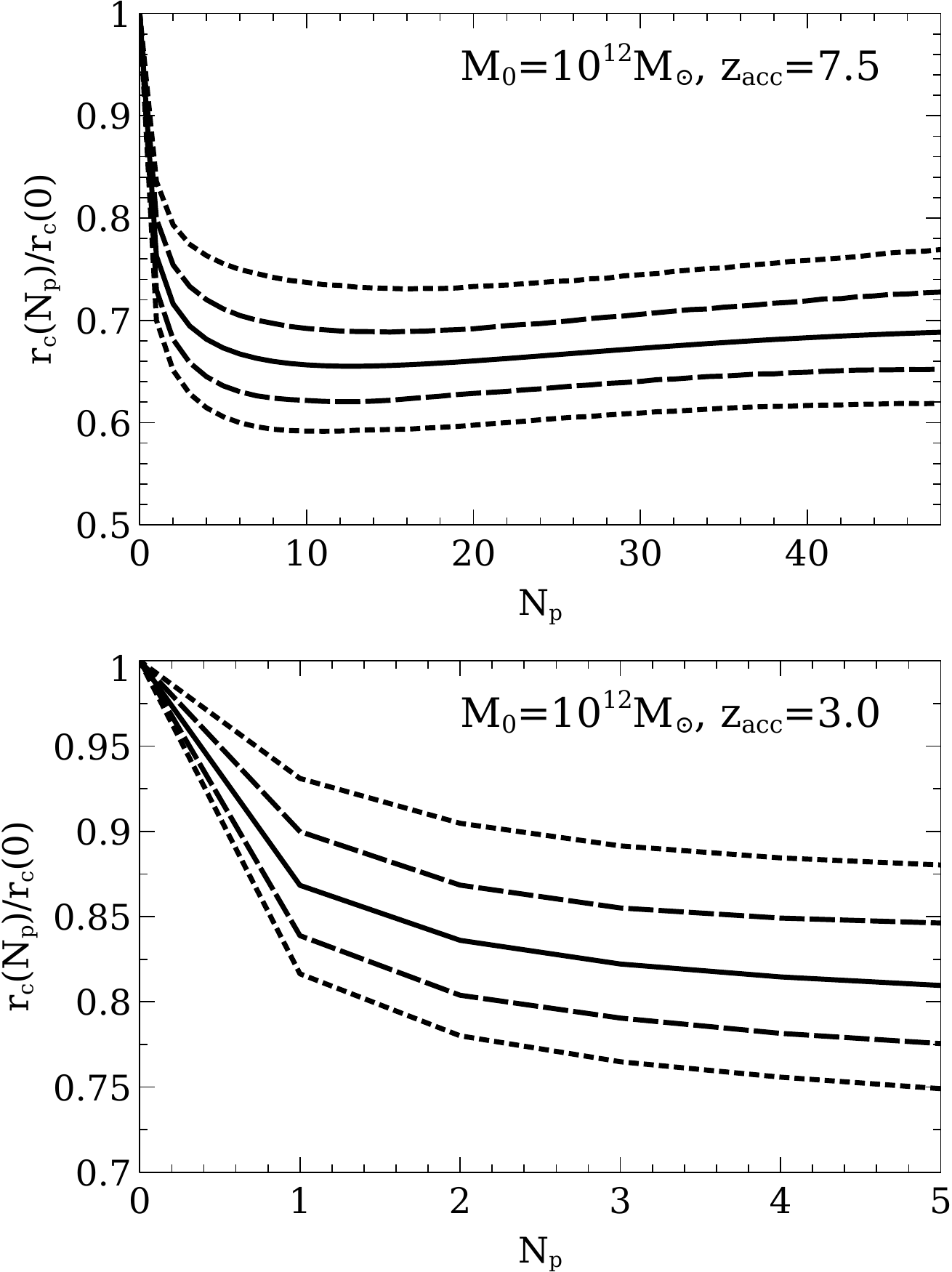}
    \end{center}
    \caption{
        Evolution of $r\sub{c}$ as a function of the number of orbital periods, $N\sub{p}$. The upper (lower) panel shows results for $M\sub{0}=10^{12}\,\msun$ and $z\sub{acc}=7.5$ (3.0). The solid line shows the mean change in $r\sub{c}$. Upper and lower dashed (dotted) lines show the 75th and 25th (90th and 10th) percentiles of the distribution. In the first few orbits, $r\sub{r}$ decreases significantly, but the decrease is stalled thereafter.
    \label{fig:rc}}
\end{figure}
In \autoref{fig:rc}, we show the evolution of $r\sub{c}$ as a function of orbital period $N\sub{p}$, for the model with $M\sub{0}=10^{12}\,\msun$. Since $r\sub{c}$ corresponds roughly to the apocentric radius, \autoref{fig:rc} indicates that the smooth growth of the host halo shrinks subhalo orbits over the first few orbital periods. The contraction in orbital radius is similar to the evolution of $J\sub{r}$, with significant change over the first few orbital periods, and little change thereafter. In the upper panel ($z\sub{acc}=7.5$), we find that the subhalo orbit actually expands slightly during the adiabatic phase ($N\sub{p} \ga 10$). The virial radius of the host halo and the apocentre of subhaloes accreted at $z\sub{acc}=7.5$ are both $\sim 5$\,kpc. The orbital expansion is driven by the decrease in the mass of the host halo at $r \la 5$\,kpc shown in \autoref{fig:profile_fiducial}. This expansion is relatively unimportant, however, compared with the orbital contraction in the early non-adiabatic phase, and the overall orbital evolution would be almost unchanged if the central density of the host halo did not decrease with time, as in \cite{Diemand2007}. The lower panel shows that for $z\sub{acc}=3.0$, $r\sub{c}$ decreases monotonically with time. At $z\sub{acc}=3.0$, the virial radius of the host halo is $\sim 50$\,kpc (\autoref{fig:profile_fiducial}) and recently accreted subhaloes have an apocentre of a similar value. In this radial range, the enclosed mass increases monotonically with time, and thus the orbit only ever contracts. We confirm that the evolution of the pericentre and apocentre resembles the $r\sub{c}$-evolution closely, i.e.~they are reduced by almost the same factor over the first few periods, while the orbital contraction then stops during the later adiabatic phase.

As indicated in \autoref{fig:pot_change_time}, the growth of the host halo potential is less adiabatic when $z\sub{acc}$ and/or $M\sub{0}$ is larger. This leads to the more significant reduction of $r\sub{c}$ (up to a factor of $\sim$1.5) in the first few orbital periods. We also find that $r\sub{c}$ is reduced by a larger factor in orbits with higher $x\sub{c,i}$ or $\eta\sub{i}$. This is because subhaloes on these orbits spend more time in the outskirts of the host halo, where the adiabatic condition is strongly broken.

\subsection{Evolution of orbital parameters}
\label{ssec:evolution_orb_params}
In predicting the orbital evolution of subhaloes in the smoothly-growing host halo potential, $L\sub{c}$ has the advantage relative to $J\sub{r}$ and $K$ that it can be evaluated at any phase of the orbit. Since the most significant part of the orbital evolution occurs rapidly (\autoref{fig:rc}), this advantage will be important for making accurate predictions. However, as shown in \autoref{fig:jr_and_proxies}, the change in $L\sub{c}$ during the evolution is greater than those in $J\sub{r}$ and $K$. If one applies a model based on $L\sub{c}$, without any further correction, the orbital evolution of subhaloes will be mispredicted. Thus, to increase the accuracy of our predictions, we introduce a correction factor. 

As shown in the numerical calculations in \autoref{ssec:how_adiabatic}, most of the change in $L\sub{c}$ occurs during the first few orbital periods. In practice, we find that the amplitude of the change in $L\sub{c}$ depends on all four possible variables, i) the accretion redshift, $z\sub{acc}$; i\hspace{-.1em}i) the final mass of the host halo, $M\sub{0}$; i\hspace{-.1em}i\hspace{-.1em}i) the orbital energy parameter at accretion, $x\sub{c,i}$; and i\hspace{-.1em}v) the orbital circularity at accretion, $\eta\sub{i}$. Motivated by this  observation, the correction factor is defined as
\begin{align}
&\frac{L\sub{c}(N\sub{\tau}, z\sub{acc}, M\sub{0}, x\sub{c,i}, \eta\sub{i})}{L\sub{c}(0, z\sub{acc}, M\sub{0}, x\sub{c,i}, \eta\sub{i})} \nonumber   \\ 
&= 1 + A \exp{(B z\sub{acc} + C M\sub{0} + D x\sub{c,i} + E \eta\sub{i})} \tanh{(F N\sub{\tau})}. 
\label{eq:lc_corrector}
\end{align}
The number of orbital periods is estimated using the smooth functional form \citep{Jiang2016},
\begin{eqnarray}
    N\sub{\tau} = \int^{z}_{z\sub{acc}} \frac{dz' (dt/dz')}{t\sub{dyn}[z', r\sub{200}(z')]}.
        \label{eq:ntau}
\end{eqnarray}
We fit the results from numerical calculations for various $M\sub{0}$ by using the {\sc curve\_fit} procedure in the {\sc scipy.optimize} module. \footnote{\url{https://www.scipy.org/}}  The fitting parameters are obtained as the averages in $k$-fold cross-validation, adopting $k=5$ \citep[e.g.,][and references therein]{Browne2000} and we derive $A=7.79 \times 10^{-4}$, $B=0.208$, $C=0.259$, $D=0.922$, $E=-0.832$ and $F=1.25$. 

\begin{figure}
    \begin{center}
        \includegraphics[width=0.45\textwidth]{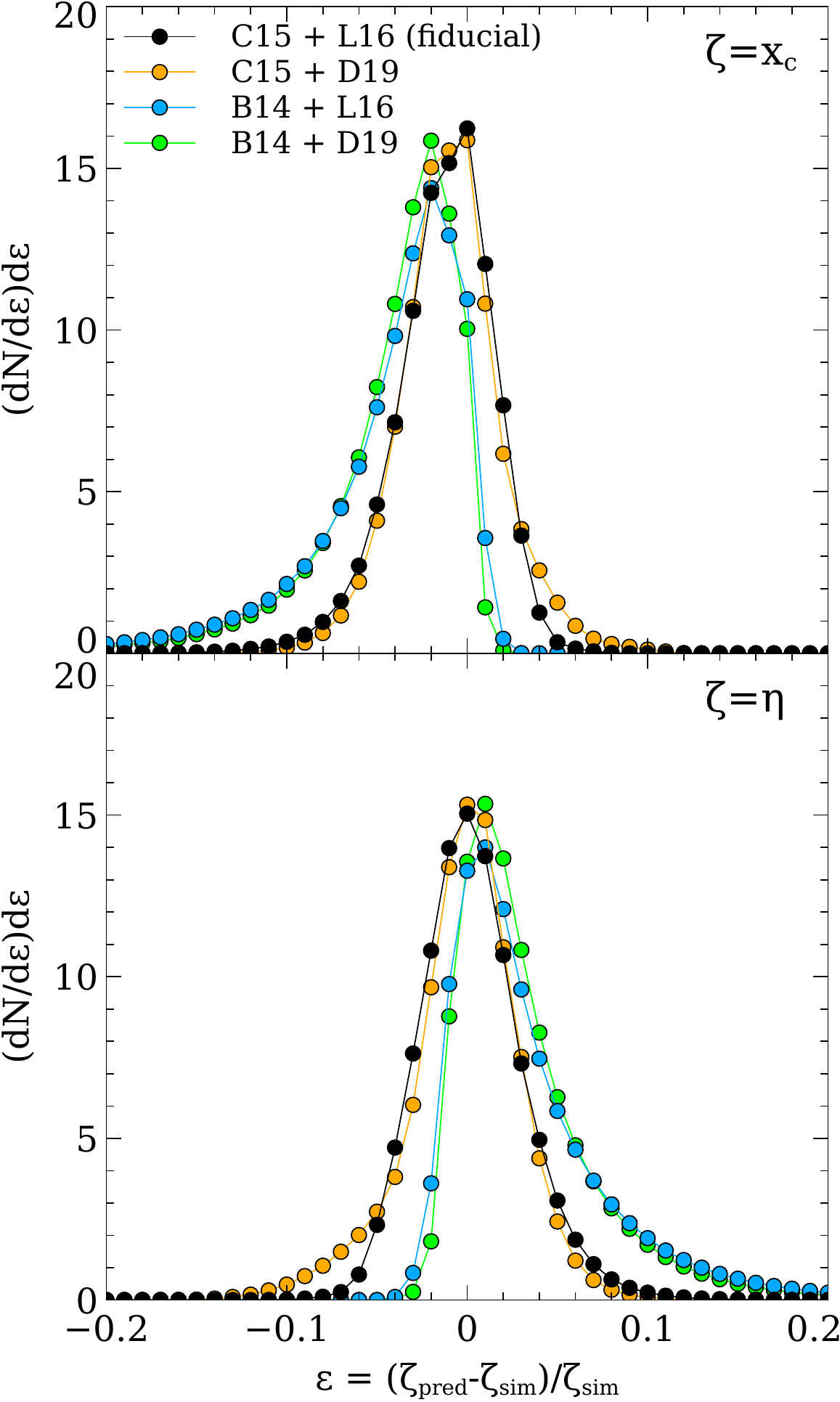}
    \end{center}
    \caption{
        Error distributions when estimating the evolved values of the orbital parameters, $x\sub{c}$ (upper) and $\eta$ (lower). Numerical calculations assuming the MAH model by \citet[][C15]{Correa2015b} and the $c(M,z)$ relation by \citet[][L16]{Ludlow2016} (black line) are used to derive the fitting parameters in \autoref{eq:lc_corrector}. Additional calculations using the MAH model by \citet[][B14]{vandenBosch2014} and/or the $c(M,z)$ relation by \citet[][D19]{Diemer2019} are also shown for reference. The general evolution of both orbital parameters is well reproduced, regardless of the detailed models used in the numerical calculation.
    \label{fig:err_distribution}}
\end{figure}
\autoref{eq:lc_corrector} enables us to accurately predict $L\sub{c}$ at a given $z$. Given the mass profile of the host halo at $z$, the predicted $L\sub{c}$ is then converted into the two orbital parameters of interest, $x\sub{c}$ and $\eta$. In \autoref{fig:err_distribution}, we show the distribution of the error in the estimate of the evolved orbital parameters. We find that they are reproduced at the five percent level. Note that data points of $z=z\sub{acc}$ are excluded from the analysis. The prediction based on \autoref{eq:lc_corrector} is applicable at any arbitrary time, and does not require the orbital integration, unlike $J\sub{r}$ or $K$. In addition, it is more accurate than an estimate of evolved orbital parameters based on $J\sub{r}$ ($K$), which can change by $\sim 10$ (25) percent in the early non-adiabatic phase (\autoref{fig:jr_and_proxies}). The fitting parameters in \autoref{eq:lc_corrector} are derived from the numerical calculations employing the MAH model by \citet[][C15]{Correa2015b} and the $c(M,z)$ relation by \citet[][L16]{Ludlow2016}. The details of the structural evolution of the host halo and the orbital evolution of subhaloes could in principle depend on this choice of models. To study the dependence, we have performed additional numerical calculations employing the MAH model by \citet[][B14]{vandenBosch2014} and/or the $c(M,z)$ relation by \citet[][D19]{Diemer2019}. While the exact values of the fitting parameters in \autoref{eq:lc_corrector} do depend on the models chosen, we find that the parameter set listed above provides excellent accuracy in describing the evolution of $x\sub{c}$ and $\eta$, even in numerical calculations with the other models.

\section{Spatial distribution of DM subhaloes}
\label{sec:spat_dist}

\begin{figure}
    \begin{center}
        \includegraphics[width=0.45\textwidth]{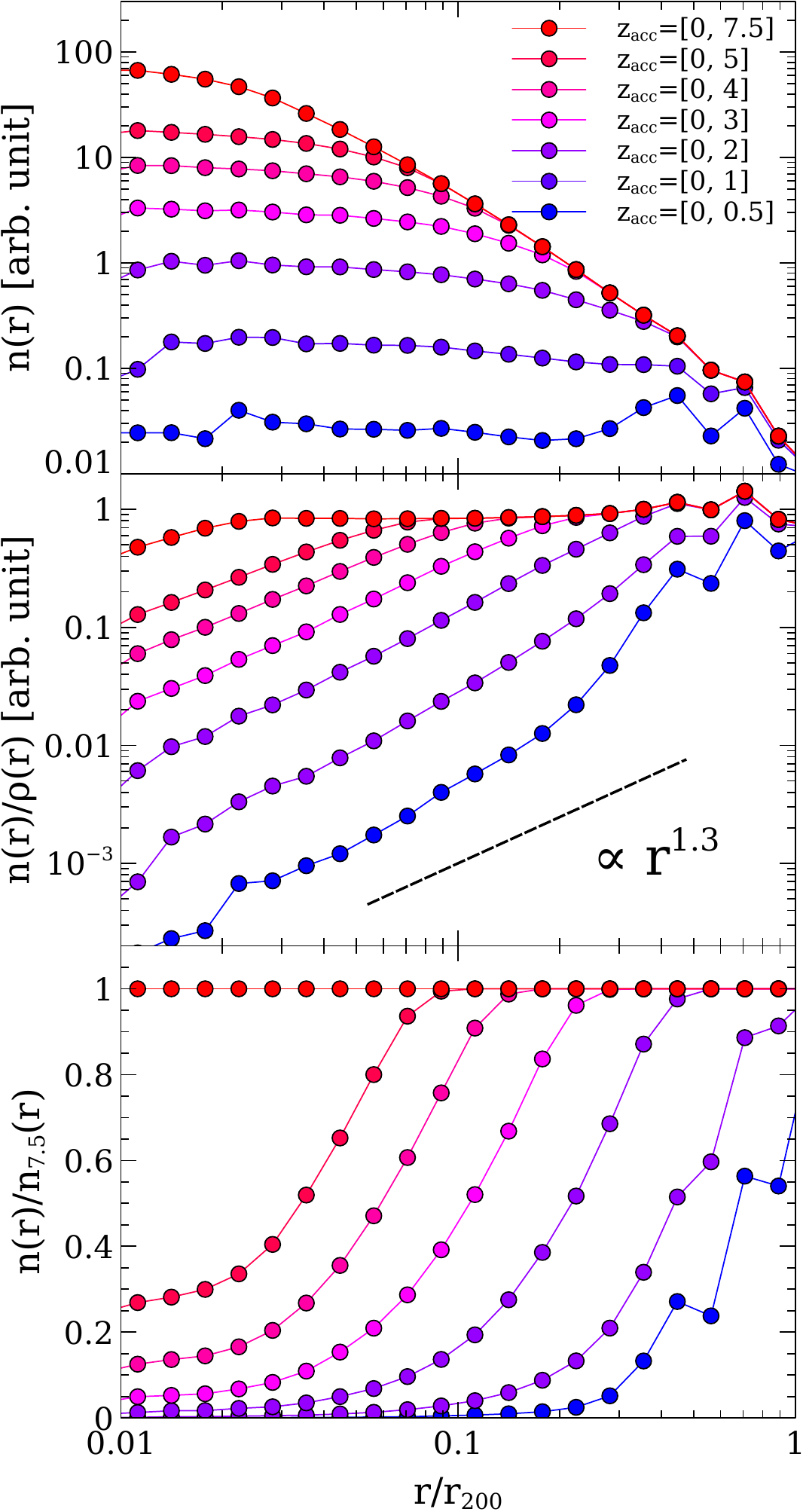}
    \end{center}
        \caption{Spatial distribution of massless particles representing subhaloes in the numerical calculation of $M\sub{0}=10^{12}\,\msun$. The snapshot at $z=0$ is used. 
        The red line shows the distribution of all particles while the other lines show the subsets selected by accretion redshift, $z\sub{acc}$. 
        ({\it Top}) Radial profile of the number density of the particles, $n(r)$. 
        ({\it Middle}) The ratio of $n(r)$ to the host halo density profile, $\rho(r)$. The dashed line is the scaling found in the Aquarius simulations \citep{Springel2008,Han2016}. 
        ({\it Bottom}) The ratio of $n(r)$ to that of $z\sub{acc}=[0, 7.5]$, $n\sub{7.5}(r)$, showing the accumulation history of subhaloes. 
        The radial bins are normalised by the virial radius of the host halo, $r\sub{200}$, and $n(r)$ is given in arbitrary units. The subhalo distribution found in the cosmological simulation matches the distribution of subhaloes accreted at $z\sub{acc} \la 3$. 
        \label{fig:subhalo_spat_dist}}
\end{figure}

The spatial distribution of DM subhaloes has been an important subject of study with cosmological $N$-body simulations \citep[][and references therein]{Ghigna2000,Diemand2004,Nagai2005,Ludlow2009,Gao2012,Hellwing2016}. These authors have shown that the radial number density profile of subhaloes within their host halo, $n(r)$, has a core of constant number density at the centre of the host halo and decreases with distance from the centre of the host halo, $r$, but that its slope is shallower than that of the host halo density profile, $\rho(r)$. Motivated by this observation, \cite{Han2016} advocated a modification of the density profile of the host halo that represents the number density of subhaloes as
\begin{eqnarray}
    n(r) \propto r^{\gamma}\rho(r),
        \label{eq:nr}
\end{eqnarray}
where $\gamma$ is a parameter controlling the significance of the modification. They found that $\gamma=1.3$ explains the subhalo distribution in the Aquarius simulations \citep{Springel2008}. \cite{Han2016} also derived the expected value of $\gamma$ analytically, based on the number of merging subhaloes, and the rate at which subsequent tidal mass-loss can lead to complete disruption of subhaloes.

Recent papers have cast doubt on the accuracy of tidal evolution for subhaloes in cosmological simulations, and thus on the actual rate of tidal disruption. For instance, \cite{vandenBosch2018a} used analytical formulae from the literature to estimate the tidal mass-loss rate due to tidal shocking and stripping, and found that neither mechanism can  explain the rate of subhalo disruption seen in cosmological simulations. \cite{vandenBosch2018b} showed that artificial subhalo disruption occurs when the force softening is inadequate, or the number of particles used to model a subhalo is too small. They concluded that many of the subhalo disruptions seen in cosmological simulations are artificial \citep[see also][]{Errani2020}. 

Motivated by this situation, we ran one more numerical calculation for a Milky Way-like host halo (final host halo mass of $M\sub{0}=10^{12}\msun$) like those in the Aquarius simulations, to study the subhalo spatial distribution free from artificial disruptions. The result from the cosmological $N$-body simulations that $n(r)$ is independent of the subhalo mass \citep{Hellwing2016} justifies the use of massless particles, since the impacts of dynamical friction and self-friction would be negligible for subhaloes with low enough masses, as shown in \autoref{ssec:mass_less_particle}. Based on \cite{Fakhouri2010}, the merger rate at $z$, $dN/dz$ scales as  
\begin{eqnarray}
    \frac{dN}{dz} \propto M\sub{200}(z)^{1.128}(1+z)^{0.0993}. 
        \label{eq:dn_dz}
\end{eqnarray}
While the halo mass definition in \cite{Fakhouri2010} is different from $M\sub{200}$ (1-3 times greater than $M\sub{200}$; \citealt{Jiang2014_mfof}), we use the scaling relation they found with $M\sub{200}$. The simple scaling works for subhaloes with low enough mass, the subject of this paper. In our additional numerical calculation, the number of subhalo massless particles accreted at $z$ is determined with \autoref{eq:dn_dz} and $N\sub{tot}=10^7$ particles accrete in total from $z=7.5$ to 0. While $N\sub{tot}$ does not correspond to the actual number of subhaloes accreted into a single host halo, we use a large number of massless particles to improve the statistics. At $z=z\sub{acc}$, massless particles are introduced at $r\sub{i} = r\sub{200}(z\sub{acc})$, with an inward velocity based on results from cosmological $N$-body simulations. We draw $x\sub{c}$ and $\eta$ by the rejection sampling from the probability distribution function (PDF) of the orbital parameters measured by \cite{Jiang2015}. Since the PDF does not strongly depend on host-to-subhalo mass ratio nor on redshift, we use the fitting result for host haloes of the virial mass of $10^{12}\,\msun$ at $z=0$ and mergers with the mass ratio of 0.0001?0.005 in the numerical calculation. Other parameters are fixed as explained in \autoref{sec:sim_model}. 

The top panel of \autoref{fig:subhalo_spat_dist} shows $n(r)$ obtained from the particle data at $z=0$. When we constrain the accretion redshift to lower $z\sub{acc}$ (bluer lines), the central number density gets lower and the size of the central core becomes larger. The orbital energy of subhaloes accreted earlier is lower, and such subhaloes live in the centre of the host halo. In the middle panel, we show the ratio of $n(r)$ to $\rho(r)$. We find that when taking subhaloes accreted at $z\sub{acc} \la 3$, the ratio $n/\rho$ shows a power-law behaviour, and that the slope is consistent with the value found by \cite{Han2016} ($\gamma=1.3$; dashed line). This result implies that the cosmological simulations can resolve only subhaloes accreted at $z\sub{acc} \la 3$, and that the profile found in cosmological simulations may be biased.\footnote{For instance, \cite{Han2016} selected subhaloes having at least 1,000 particles in their analysis.} We show in the bottom panel the ratio of $n(r)$ to that of $z\sub{acc}=[0, 7.5]$, $n\sub{7.5}(r)$, indicating that subhaloes accreted earlier are located close to the centre, while ones accreted later dominate the outskirts of the host halo. This panel implies that inferring the accretion epoch of subhaloes from their position is possible \citep[see also][]{Oman2013}. For instance, looking at the pink line ($z\sub{acc}=[0,3]$), the ratio exceeds 0.5 at $r/r\sub{200} \sim 0.1$. Subhaloes at $r/r\sub{200} < 0.1$ ($r/r\sub{200} > 0.1$) are inferred to have been accreted at $z\sub{acc} > 3$ ($z\sub{acc} < 3$). 

\autoref{fig:subhalo_spat_dist} reveals another interesting implication. When all particles are used (red line), $n(r)$ is almost cuspy at the centre (top panel). Comparing $n(r)$ to $\rho(r)$, the subhalo distribution has almost the same radial dependence as the density profile of the host halo (middle panel). This was in fact implied by \cite{Han2016}. They traced the position of unresolved (disrupted) subhaloes virtually, by tracking the most bound particle from each disrupted subhalo, and found that their spatial distribution is similar to the density profile of the host halo in a broad radial range ($0.01 \la r/r\sub{200} \la 1$). A fraction of the traced particles sank to the centre further by the impact of dynamical friction and formed a steeper cusp. We will study this point in more detail in the companion paper (Ogiya et al., in prep.).

\section{Mass evolution of the possible progenitor of the DM deficit galaxy}
\label{sec:df2}

\begin{table*}
    \begin{center}
        \caption{Summary of model parameters describing the mass evolution of the DF2 progenitor. 
                 Description of each column: (1) accretion redshift. (2) lookback time. (3) host halo mass. (4) concentration of the host halo. (5) subhalo mass. (6) concentration of the subhalo. (7) probability of orbital parameter sets reproducing the DF2 mass criterion, when neglecting the orbit contraction due to the smooth growth of the host halo. (8) probability of orbital parameter sets reproducing the DF2 mass criterion, when taking the orbit contraction into account. Since the host halo mass and structure at $z=0$ are required in the analysis, they are shown in the first row.
        \label{tab:df2}}
            \begin{tabular}{cccccccc}
                \hline 
                (1)          & (2)                 & (3)                     & (4)            & (5)                    & (6)        & (7)                   & (8)    \\
                $z\sub{acc}$ & $t\sub{lb}$ [Gyr]   & $M\sub{h}$ [$\msun$]    & $c\sub{h}$     & $M\sub{s}$ [$\msun$]   & $c\sub{s}$ & $P\sub{no-cont}$      & $P\sub{with-cont}$    \\
                \hline 
                0.0          & --                  & $6.2 \times 10^{12}$    & 7.3            & --                         & --       & --                    & -- \\
                1.0          & 7.9                 & $3.1 \times 10^{12}$    & 5.6            & $5.9 \times 10^{10}$     & 7.7      & 0.0                   & 0.0           \\
                1.5          & 9.5                 & $2.1 \times 10^{12}$    & 5.0            & $6.0 \times 10^{10}$     & 6.6      & $3.6 \times 10^{-5}$  & $2.8 \times 10^{-4}$ \\
                2.0          & 10.5                & $1.4 \times 10^{12}$    & 4.6            & $6.0 \times 10^{10}$     & 5.7      & $3.3 \times 10^{-5}$  & $3.7 \times 10^{-4}$ \\
                \hline
            \end{tabular}
    \end{center}
\end{table*}

\begin{figure}
    \begin{center}
        \includegraphics[width=0.45\textwidth]{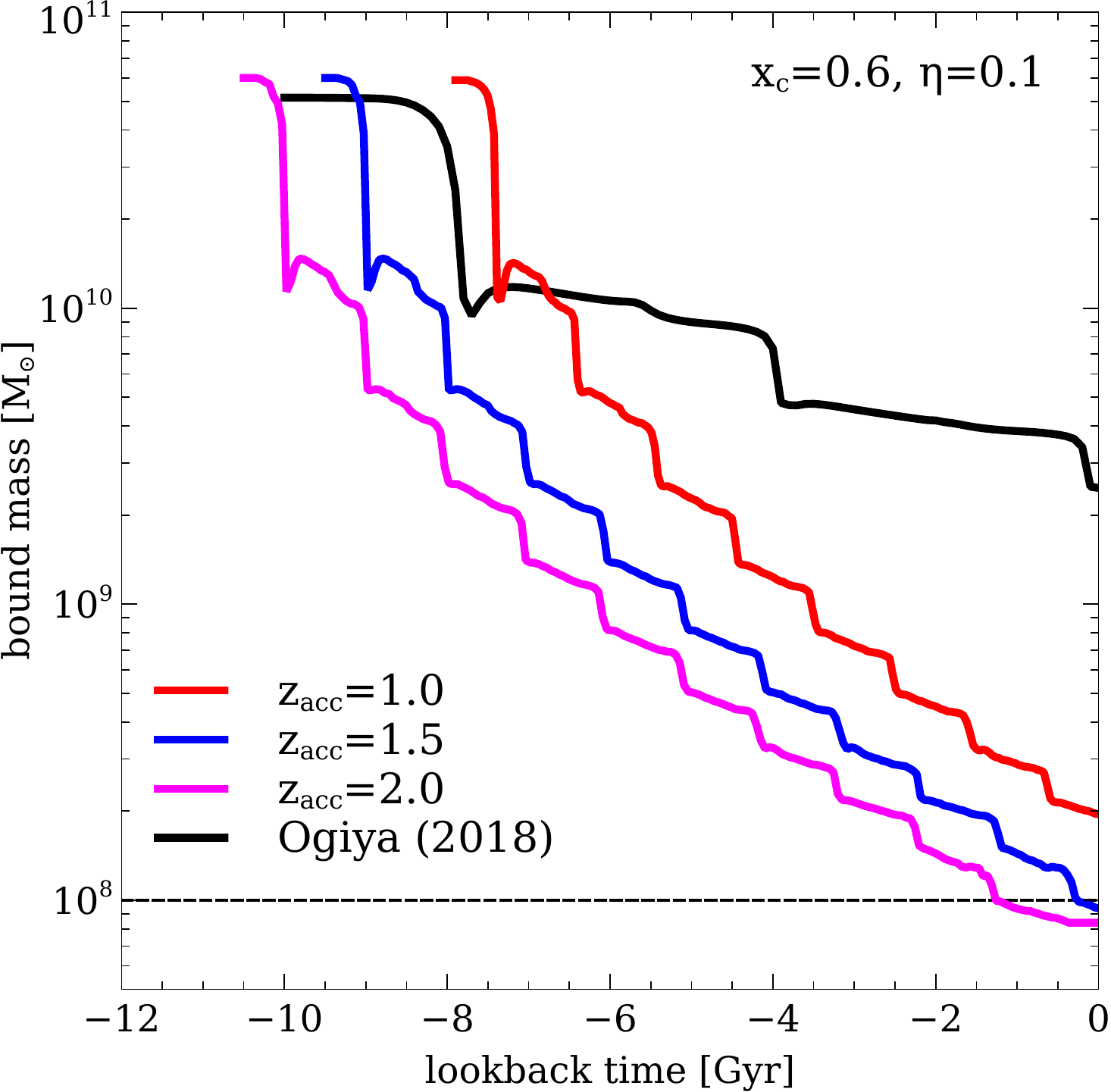}
    \end{center}
        \caption{An example of the bound mass evolution of the DF2 progenitor models. Red, blue and magenta lines show the cases in which the progenitor accreted into the host system (main progenitor of NGC1052) at $z\sub{acc}=1.0$, 1.5 and 2.0. The orbit contraction due to the smooth growth of the host halo is neglected. For comparison, the DM halo mass evolution from the simulation of the cuspy halo model by \citet{Ogiya2018} is shown as the black line. The pair of dimensionless parameters characterising the merger orbit is $x\sub{c}=0.6$ and $\eta=0.1$. Horizontal dashed line is the upper DM mass limit for DF2, inferred by \citet{vanDokkum2018}.
        \label{fig:mb}}
\end{figure}

Subhaloes accreted at higher $z\sub{acc}$ have lower orbital energies (i.e.~smaller $x\sub{c}$) at a given $z$. They have smaller pericentres and feel stronger tidal forces from their host halo than those accreted later. In addition, their orbital period is shorter, and thus the number of pericentric passages in a fixed interval of time is larger. Thus subhaloes accreted earlier will experience more significant tidal stripping in their evolution. 

We here consider the tidal evolution of an ultra diffuse galaxy, NGC1052-DF2 (hereafter, DF2), in the group centred on the large elliptical galaxy NGC1052. Recently, observations \citep{vanDokkum2018,Wasserman2018,Danieli2019} have inferred that the DM mass contained in DF2 is several hundred times smaller than expected from the empirical models of galaxy formation and evolution \citep{Moster2018,Behroozi2019}. While there is considerable debate on this interpretation, e.g., discussion of the overall statistical confidence due to the small number of kinematic tracers \citep{Martin2018,Laporte2019}, the details of the data processing \citep{Hayashi2018,Trujillo2019}, and the consistency with the orbital decay timescale due to dynamical friction \citep{Nusser2018,DuttaChowdhury2019}, considering the possible formation processes of such extreme galaxies is an interesting and important step in advancing our understanding of galaxy formation and evolution \citep{Leigh2019,Sales2019}. For instance, \cite{Ogiya2018} showed that DF2 could be a result of a violent tidal stripping event, but that a cored DM density profile, rather than a cuspy profile like the NFW density profile (\autoref{eq:nfw}), is necessary to reproduce the observations \citep[see also][]{Penarrubia2010,Yang2020}.

A caveat to the proposal of \cite{Ogiya2018} is that the structure of galaxies and the orbital parameters prior to tidal interactions were based on observations and empirical relations at $z=0$. Since DF2 is a member of the galaxy group, the progenitor of DF2 must have accreted at $z\sub{acc}>0$ and the modelling of tidal stripping will be improved by taking the subsequent evolution of the subhalo orbit into account. The empirical model of galaxy formation and evolution by \cite{Behroozi2019} showed that the peak halo mass (i.e.~mass prior to mergers into larger systems) is a good indicator for the stellar mass of satellite galaxies. Inversely, we can estimate the subhalo mass at $z\sub{acc}$, $M\sub{s}$, from the current stellar mass of DF2, $\sim 2 \times 10^8 \msun$ \citep{vanDokkum2018}. The empirical model also predicts the mass growth history of the halo surrounding the host galaxy, NGC1052, given its current stellar mass, $\sim 10^{11} \msun$ \citep{Forbes2017}. Once the virial masses of the host- and sub-haloes at $z\sub{acc}$ are obtained, we use the $c(M,z)$ relation by \cite{Ludlow2016} to determine the structure of the two merging systems, assuming NFW density profiles. \autoref{tab:df2} lists the parameter sets obtained for $z\sub{acc}=1.0$, 1,5 and 2.0. 

We re-examine the tidal stripping model for a cuspy NFW density profile, using the DASH library of idealised $N$-body simulations of minor halo mergers.\footnote{\url{https://cosmo.oca.eu/dash/}} \cite{Ogiya2019} ran more than 2,000 high-resolution simulations covering a broad range in the parameters determining the structure of the haloes and the orbit of the subhalo. They then trained a non-parametric machine learning model to reproduce the mass evolution of subhaloes in the tidal field of the host halo potential, based on the random forest algorithm \citep{Breiman2001} as implemented in {\sc scikit-learn} \citep{Scikit-learn}\footnote{\url{https://scikit-learn.org}}. Relative to the simulations, the final predictions of the model are accurate at the 0.1\,dex level. Although the DASH simulations consider mergers between pure dark matter structures, they should give a reasonable indication of the overall behaviour. The DASH simulations assume that both the host- and sub-haloes have NFW density profiles prior to the merger, and that the ratio of the host halo to subhalo mass, $f \equiv M\sub{h}/M\sub{s}$, is large. In such mergers, we can safely neglect the impacts of dynamical friction and self-friction. While the DASH library is only formally applicable to mergers with $f \ga 100$, we use it in modelling the tidal evolution of DF2 where $f \sim 20-50$ (see \autoref{tab:df2}). Note that dynamical friction would not be completely negligible in those cases (\autoref{fig:fcrit}). We neglect the growth of the host halo potential, i.e.~the structure of the host halo is fixed from $z=z\sub{acc}$ to 0, and orbital decay due to dynamical friction and self-friction, as assumed in the simulations performed by \cite{Ogiya2019}.

To study the importance of orbital contraction driven by the smooth growth of the host halo, two cases, with and without orbital contraction, are considered for each $z\sub{acc}$. In the models neglecting orbital contraction, the orbital parameters are unchanged since accretion. In the other models, we virtually take orbital contraction into account by using \autoref{eq:lc_corrector}. While the orbital parameters are predicted for $z=0$, they are applied at $z=z\sub{acc}$. The orbital circularity, $\eta$, is used as it is predicted, and $x\sub{c}$ is scaled by multiplying the factor of $r\sub{200}(z=0)/r\sub{200}(z\sub{acc})$ to remove the effect of the growth of the virial radius from the $x\sub{c}$-evolution.\footnote{Because of the potential growth, the same values of $r\sub{c}$ at two different redshifts do not correspond to the same orbital energy. This is neglected in the model for simplicity.} The orbit is assumed to shrink instantaneously at $z\sub{acc}$, while the actually orbit will in fact shrink more gradually (\autoref{fig:rc}). Thus, this treatment provides an upper limit on the estimated strength of the effect of the orbit contraction.

\citet{vanDokkum2018} inferred the enclosed DM mass within 3.1 and 7.6\,kpc from the centre of DF2. On the other hand, the machine learning model predicts the mass gravitationally bound to the subhalo at given time measured from the beginning of accretion. The comparison is nonetheless justified, for two reasons. First, the process of tidal stripping removes the mass preferentially from the outskirts of the subhalo, and hardly changes its central density structure \citep[e.g.,][]{Hayashi2003,Penarrubia2010,Ogiya2019}. Second, the radius where the enclosed mass goes below the upper mass limit in the subhalo mass profile prior to accretion is $\sim 0.4-0.5$\,kpc, smaller than 3.1\,kpc in all models. Thus, if the estimated bound mass is below the upper mass limit, the enclosed mass criteria would be satisfied. 
\footnote{An additional justification is provided by the model for the density profile of tidally stripped subhaloes by \cite{Green2019} that requires the subhalo concentration and the ratio of the bound mass of the subhalo at given time to the subhalo mass prior to the merger. We use the machine learning model to compute the latter and derive the enclosed mass within 3.1\,kpc from the centre of the DF2 progenitor by the numerical integration. While the predicted bound mass is larger than the enclosed mass, the difference between them gets smaller when considering more significant mass-loss events and is reduced to $\sim$\,10 percent in the events in which the bound mass goes below the upper mass limit of DF2. Using the enclosed mass instead of the bound mass in the analysis, the probabilities of orbital parameter sets reproducing the DF2 mass criterion are the same as shown in \autoref{tab:df2}.}

In \autoref{fig:mb}, we show an example of the expected bound mass evolution of the halo surrounding the DF2 progenitor accreted at $z\sub{acc}=1.0$ (red), 1.5 (blue) and 2.0 (magenta). For comparison, the black line shows the result from the simulation of the cuspy density profile model performed by \cite{Ogiya2018}. The case of a tightly bound ($x\sub{c}=0.6$) radial ($\eta=0.1$) orbit is shown. These orbital parameters are the same as employed in \cite{Ogiya2018}. As expected, the number of pericentric passages is larger (8-10) than that seen in \cite{Ogiya2018}, 3, and the bound mass goes below the upper mass limit inferred by \citet[][horizontal dashed line]{vanDokkum2018} in the models with $z\sub{acc}=1.5$ and 2.0. 

We test if the bound mass goes below the upper mass limit by the present time in the two-dimensional space of the orbital parameters at accretion, $x\sub{c,i}=[0.5,2.0]$ and $\eta\sub{i}=[0.01:0.99]$ with the interval of $\Delta x\sub{c,i} = \Delta \eta\sub{i} = 0.01$ (i.e.~14,949 pairs of the orbital parameters) for $z\sub{acc}=1.0$, 1.5 and 2.0, and weight each model based on the PDF of \cite{Jiang2015}. We use the fitting parameters for host haloes of virial mass of $10^{13}\,\msun$ at $z=0$ and mergers with the mass ratio of 0.005-0.05 that are consistent with the halo models in the analysis. The probabilities of the orbital parameter sets in the two-dimensional space of $x\sub{c,i}$ and $\eta\sub{i}$ satisfying the mass limit when neglecting and when considering orbital contraction are shown in the seventh and eighth columns in \autoref{tab:df2}. In the model taking orbital contraction into account, we set $x\sub{c}=0.5$ when $x\sub{c}<0.5$ is predicted because of limitations of the machine learning predictions for mass-loss. We find that DF2 is not an impossibly rare object, even if its DM halo is cuspy, provided it was accreted early enough ($z\sub{acc} \ga 1.5$), . While the probability is increased by a factor of $\sim 10$ when orbital contraction is considered, it remains fairly low, however, and thus DF-2 should be considered very uncommon at best.

A follow-up observational paper reported another DM deficit galaxy in the same galaxy group, NGC1052-DF4 \citep[][but see also \citealt{Monelli2019}]{vanDokkum2019}. The probabilities we have obtained suggest that tidal stripping is very unlikely to explain the existence of two or more DF-deficit galaxies in a single galaxy group. They would increase, however, when taking into account other mechanisms that could shrink the orbit of DF2, such as dynamical friction and self-friction. The probability would increase further if the DM halo of DF2 has a cored density profile. A more detailed study is needed to reach a firm conclusion regarding the formation of these mysterious objects.

\section{Summary}
\label{sec:summary}
Subhaloes are a promising probe of the nature of DM. Astrophysical observations can constrain their masses and spatial distribution, but accurate theoretical predictions of their properties are needed to support these efforts. The mass evolution and spatial distribution of subhaloes within the host halo depend strongly on their orbital parameters, which in turn are influenced by several different processes. Some of these (e.g.~dynamical friction and self-friction) are negligible for all but the most massive subhaloes, while others (e.g.~violent relaxation driven by major mergers) are hard to model analytically. In this paper we have considered two other processes, interactions between subhaloes, and the smooth component of mass growth of the host halo, that occurs even in realistic systems formed by hierarchical merging. Based on a simple model of subhalo-subhalo interactions, we find that these can erase the memory of initial subhalo orbital parameters by the present day, if systems are accreted at $z\sub{acc} \ga 5$, while large changes due to strong encounters are rare and thus unimportant. For systems accreted at later times, the smooth change in the background potential is the main effect driving orbital evolution.

To isolate this effect from the other mechanisms, we use numerical calculations with massless particles that represent subhaloes orbiting in the smoothly evolving potential of a host halo. Since the subhalo particles do not have mass, they do not feel drag forces, and the interactions between them are neglected. We find that the radial action of subhalo orbits, $J\sub{r}$, which is a conserved quantity in spherical systems evolving adiabatically, decreases by $\sim$\,10 percent soon after accretion into the host halo, and stays almost constant thereafter. The non-conservation indicates that the smooth mass growth of the host halo is not adiabatic for subhalo orbits, and as a result they shrink by a factor of $\sim$\,1.5. During this evolution, the change in the proxies for $J\sub{r}$, $K$ and $L\sub{c}$, is larger than that in $J\sub{r}$, but the overall form of the evolution is similar to that of $J\sub{r}$. We introduce an analytic model for the evolution of the orbital parameters based on $L\sub{c}$, since it can be evaluated at an arbitrary phase of the orbit.

Our model based on the corrected $L\sub{c}$ should accurately describe the orbital evolution of subhaloes whose mass is small enough compared to that of the host, that dynamical friction and self-friction are negligible. For more massive subhaloes, these mechanisms should be properly taken into account. While we have assumed the spherical host halo potential, DM haloes in cosmological simulations are in fact triaxial \citep[][and references therein]{Jing2002,Kuhlen2007,Vera-Ciro2014}. The modelling for spherical systems would be applicable for triaxial systems with some modification of the proxy for $J\sub{r}$ \citep[e.g.,][]{Lithwick2011}. The proper modification would depend on the triaxiality of the host haloes and further investigations are needed for accurate predictions of the subhalo orbital evolution.  Relaxing the simplifying assumptions made in this paper would lead to a more realistic model of orbital evolution, although it would make analytical investigations far more complicated. To understand the statistical evolution of subhalo orbits, analysing the data from fully realistic cosmological simulations would be a promising avenue.

We study the spatial distribution of subhaloes in a Milky Way-sized host halo and find that in current high-resolution cosmological simulations, the dominant fraction of subhaloes surviving at $z=0$ may have been accreted at $z\sub{acc} \la 3$, while those accreted earlier are unresolved. We also consider the implications of our numerical calculations for the mass evolution of a DM-deficient galaxy, DF2. If the progenitor of DF2 was accreted into the host galaxy at high enough redshift, $z\sub{acc} \ga 1.5$, with the orbital parameters of $x\sub{c} \sim 0.6$ and $\eta \sim 0.1$, tidal stripping by the host galaxy potential can reproduce the upper mass limit inferred from observations. The required orbital parameters are in the tail of the PDF, however, making this scenario somewhat unlikely, even if the orbit contraction driven by the smooth growth of the host halo is taken into account. Additional factors increasing the overall mass-loss efficiency, such as a further reduction of the orbital size by dynamical friction and self-friction, or increase susceptibility to mass-loss due to a cored DM density profile, would be needed to explain the existence of two or more DM-deficient galaxies in the galaxy group.

\section*{Acknowledgements}
We thank the anonymous referee and Neal Dalal for providing the insightful comments that greatly improved the paper. We are grateful to the developers of {\sc SciPy} and {\sc scikit-learn} for making their code publicly available. A part of numerical calculations was performed on the Graham cluster operated by Compute Canada (\url{www.computecanada.ca}). JET and MJH acknowledge financial support from NSERC Canada, through Discovery Grants. 

\section*{Data Availability}
The data underlying this article will be shared on reasonable request to the corresponding author.



\bibliographystyle{mnras}
\bibliography{go} 



\appendix
\section{Toy model for subhalo-subhalo interactions}
\label{app:model_subhalo_subhalo}

This appendix provides a simple estimate of the relative importance of subhalo-subhalo interactions in the orbital evolution of subhaloes (the final results of the analysis are presented in \autoref{sssec:subhalo_subhalo}). The model is a modified version of the argument presented in \S\,1.2.1 of \cite{Binney2008}. 

In this model for interactions, subhaloes are assumed to be point masses. Suppose that a subhalo moves on a straight path passing through the centre of the host halo and passes by another subhalo with a mass of $M\sub{s}$ (the `perturber'). For simplicity, during the interaction, the perturber is fixed and the relative velocity between them is a constant, $v$. We denote the impact parameter (the perpendicular distance between the path and the perturber) as $b$. The velocity perturbation in the perpendicular direction in an interaction is 
\begin{eqnarray}
    dv = 2GM\sub{s}/bv. 
        \label{eq:dv}
\end{eqnarray}
When $b$ is smaller than 
\begin{eqnarray}
    b\sub{90} \equiv 2GM\sub{s}/v^2,
        \label{eq:b90}
\end{eqnarray}
$dv$ can be greater than $v$, in which case and the orbit of the `subject' subhalo will be deflected by more than 90 degrees. We refer to such interactions as close or strong encounters. Even if $b > b\sub{90}$, cumulative impacts of weak interactions with multiple perturbers can alter the orbit of the subject subhalo. Supposing that perturbers are isotropically distributed in the host halo, $\Delta v \equiv \sum dv = 0$, while $\Delta v^2 \equiv \sum dv^2 = dN \times dv^2 > 0$. Here $dN$ represents the number of perturbers in $[b : b+db]$ and $dN = 2 \pi b \Sigma(b) db$. The column number density of the subhalo is derived by the integration,
\begin{eqnarray}
    \Sigma(b) = \int^{\sqrt{r\sub{200}^2-b^2}}_{0} 2 n\bigl( \sqrt{z^2+b^2} \bigr)dz,
        \label{eq:Sigma_b}
\end{eqnarray}
where $n(r)$ is the number density profile of subhaloes in the host halo. 

Host haloes contain multiple subhalo populations in terms of their masses. The efficiency of perturbers in altering the subject subhalo's orbit will depend on their mass, $M\sub{s}$. When $M\sub{s}$ is larger, the impact of a single interaction is larger. For instance, $dv^2$ is proportional to $M\sub{s}^2$. On the other hand, subhaloes with smaller masses are more abundant than those with larger masses \citep[e.g.,][]{Giocoli2008,Springel2008,Jiang2016}. According to cosmological $N$-body simulations, the subhalo mass function roughly scales as $dN/d\ln{(M\sub{s}/M\sub{h})} \propto (M\sub{s}/M\sub{h})^{-1}$ in the limit of $M\sub{s}/M\sub{h} \ll 1$, where $M\sub{h}$ is the host halo mass. Therefore massive perturbers are the main contributor in altering the subject subhalo's orbit in a cumulative fashion. They also more efficiently alter the subject subhalo's orbit in the close encounter channel, as indicated by \autoref{eq:b90}. 

Given this argument, we focus on encounters with massive subhaloes. As discussed in detail in \autoref{sec:spat_dist}, \autoref{eq:nr} models the spatial distribution of recently accreted subhaloes within the host halo. Such subhaloes are more massive than those accreted earlier because of the nature of the hierarchical structure formation (larger structures are formed, and merge, later) and because they have experienced less tidal mass-loss. We assume that all subhalo populations follow the distribution of \autoref{eq:nr} with $\gamma=1.3$, and compute the column number density of subhaloes of all populations (\autoref{eq:Sigma_b}). Then the column number density of subhaloes with $M\sub{s}$ is given by
\begin{eqnarray}
    \Sigma(b, M\sub{s}) = \Sigma(b) N(M\sub{s})/N\sub{sub,tot},
        \label{eq:Sigma_b_ms}
\end{eqnarray}
where $N(M\sub{s})$ and $N\sub{sub,tot}$ are the number of subhaloes with $M\sub{s}$ and the total number of subhaloes in the host halo, respectively. We estimate $N(M\sub{s})$ and $N\sub{sub,tot}$ for a host halo with given $M\sub{0}$ at given redshift assuming the model for the MAH of DM haloes by \citet[][see \autoref{ssec:host_growth} for details]{Correa2015b} and the formulation for the subhalo mass function by \cite{Jiang2016}.\footnote{They referred to it as the `evolved subhalo mass function'.}

The square of the relative velocity between the subject subhalo and the perturber, $v^2$, appears in computing $dv^2$ and $b\sub{90}$. For simplicity, the subject subhalo is assumed to be on a radial orbit passing through the centre of the host halo and drag forces (dynamical friction and self-friction) and the mass growth of the host halo are neglected. Given that the specific orbital energy of the subject subhalo is $E=\Phi(r\sub{200})$ where $\Phi(r)$ is the gravitational potential of the host halo, the subject subhalo oscillates in the range of $X=[-r\sub{200}(z\sub{acc}) : r\sub{200}(z\sub{acc})]$ while $Y=Z=0$. We define $v^2$ as 
\begin{eqnarray}
    v^2 = \int^{r\sub{200}}_{-r\sub{200}} 2[E-\Phi(X)] \mathcal{N}(X) dX \biggl /  \int^{r\sub{200}}_{-r\sub{200}} \mathcal{N}(X) dX.
        \label{eq:v2_weighted}
\end{eqnarray}
Here, $\mathcal{N}(X)$ is computed by
\begin{eqnarray}
    \mathcal{N}(X) = \int^{\sqrt{r\sub{200}^2-X^2}}_{0} 2 \pi R n\bigl( \sqrt{R^2+X^2} \bigr) dR.
        \label{eq:mathcalN}
\end{eqnarray}
Note that in \autoref{eq:v2_weighted} the local velocity squared of the subject subhalo is weighted with the number of perturbers at $[X : X+dX]$, $\mathcal{N}(X)dX$, so that \autoref{eq:v2_weighted} corresponds to the averaged velocity squared of the subject subhalo at the the closest approach to the perturbers. This treatment is justified by the fact that the velocity perturbation of single encounters (\autoref{eq:dv}) corresponds to the product of the acceleration at the closest approach, $GM\sub{s}/b^2$, and the time duration of the interaction, $2b/v$ \citep{Binney2008}. 

Perturbers are counted as close or weak encounters based on \autoref{eq:b90}, i.e.~interactions with $b \leq b\sub{90}$ are close encounters while those with $b > b\sub{90}$ are weak encounters. Perturbers located at $\sqrt{Y^2+Z^2}/r\sub{200}=[10^{-3}:1]$ with a mass of $M\sub{s}/M\sub{h}=[10^{-6}:1]$ are considered. While \cite{Jiang2016} adopted the definition of virial overdensity by \cite{Bryan1998}, we define $M\sub{h} \equiv M\sub{200}(z\sub{acc})$, i.e., $\Delta\sub{vir}(z\sub{acc})=200$. The results of the analysis are insensitive to these parameters. We assume that the distribution and the mass function of subhaloes within $r\sub{200}(z\sub{acc})$ are unevolved from $z=z\sub{acc}$ to 0. The number of close encounters and $\Delta v^2$ are counted in a single crossing of the subject subhalo. We multiply them by the ratio of the lookback time to $z\sub{acc}$ to the crossing time of the subject subhalo, defined as twice the dynamical time of the host halo at $z\sub{acc}$,
\begin{eqnarray}
    t\sub{cross}(z\sub{acc}) \equiv 2t\sub{dyn}[z\sub{acc}, r\sub{200}(z\sub{acc})] = \sqrt{\frac{3 \pi}{800 G \rho\sub{crit}(z\sub{acc})}}.
        \label{eq:tcross}
\end{eqnarray}

\section{Evolution of host halo profiles}
\label{app:profile_evolution}

\begin{figure*}
    \begin{center}
        \includegraphics[width=0.90\textwidth]{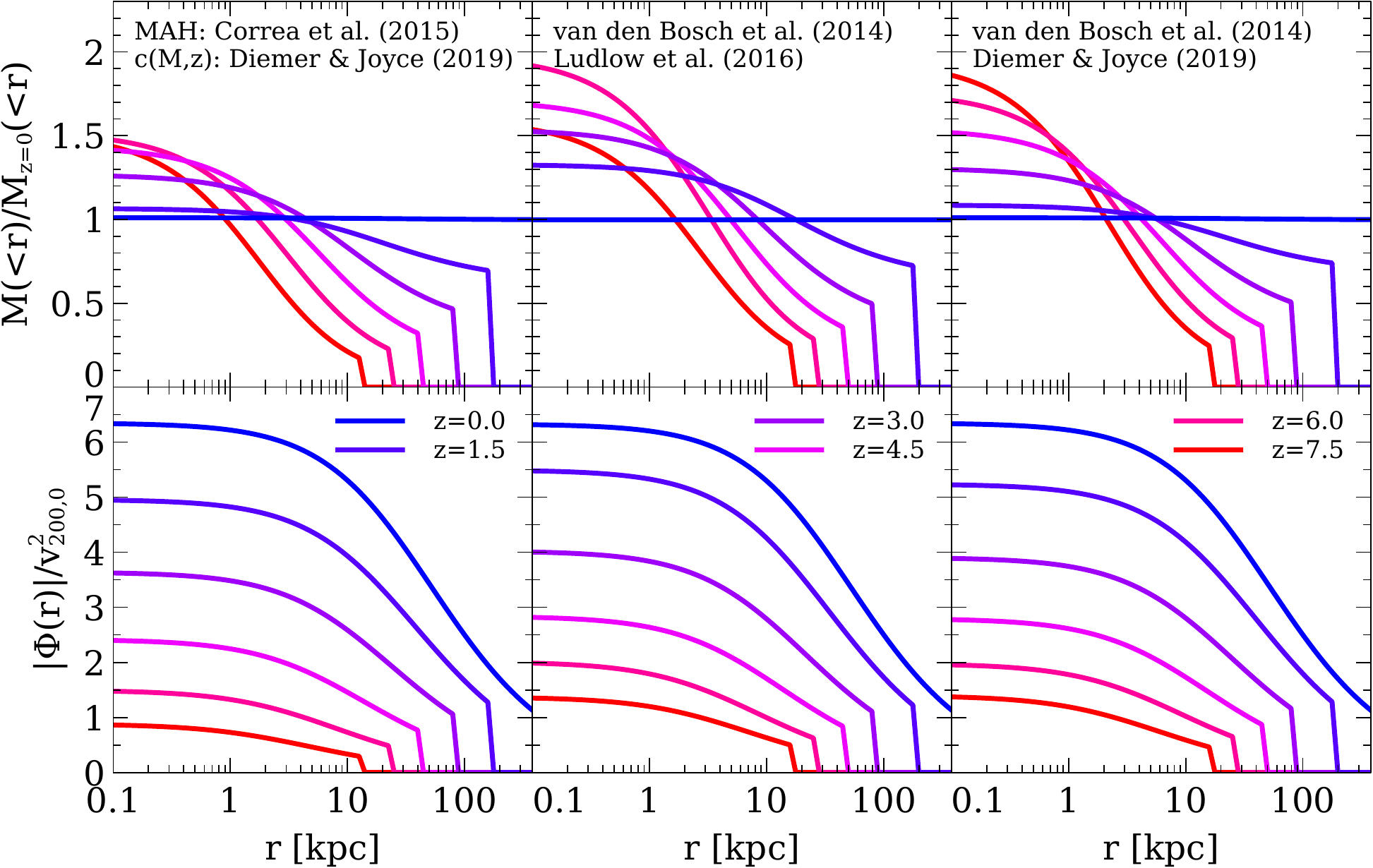}
    \end{center}
    \caption{
        Same as \autoref{fig:profile_fiducial} but varying the MAH model and/or the $c(M,z)$ relation, as indicated in the upper panels. The fiducial case shown in \autoref{fig:profile_fiducial} employs the MAH model of \citet{Correa2015b} and the $c(M,z)$ relation of \citet{Ludlow2016}. 
        ({\it Upper}) The enclosed mass profile at given $z$, $M(<r)$, scaled with that derived in the fiducial case at $z=0$, $M\sub{z=0}(<r)$. 
        ({\it Lower}) The potential profile at given $z$, $\Phi(<r)$, scaled with the virial velocity squared at $z=0$, $v\sub{200,0}^2 \equiv GM\sub{0}/r\sub{200}(z=0)$. 
        The radial bins are given in (fixed) physical kpc. The profiles at $z$ are computed in the range of $r=[0.1\,{\rm kpc}, 2\,r\sub{200}(z)]$ where $r\sub{200}(z)$ is the virial radius of the halo at $z$. 
        The potential profiles are insensitive to the choice of the MAH model or the $c(M,z)$ relation. While the details of the evolution of the central mass structure depends on the choices, the same trend, i.e., that the central density decreases with time at $z \la 6$, is obtained in all cases. The growth of the halo outskirts is similar in all cases.
    \label{fig:profile_other_models}}
\end{figure*}

In this appendix, we see how the enclosed mass and potential profiles, $M(<r)$ and $\Phi(r)$, depend on the MAH model and the $c(M,z)$ relation. The fiducial case employs the MAH model by \cite{Correa2015b} and the $c(M,z)$ relation by \cite{Ludlow2016} to specify the internal structure of the host halo at a given redshift $z$, and the radial profiles are shown in \autoref{fig:profile_fiducial}.  \autoref{fig:profile_other_models} shows the evolution of the enclosed mass and potential profiles obtained by varying the MAH model (\citealt{Correa2015b} or \citealt{vandenBosch2014}) and/or the $c(M,z)$ relation (\citealt{Ludlow2016} or \citealt{Diemer2019}), as indicated in the upper panels. We find that the derived profiles are consistent with the fiducial case; i) the mass at the halo outskirts and the virial radius of the host halo increases, while the central density decreases with time at $z \la 6$. i\hspace{-.1em}i) the potential in the centre of the halo deepens relative to the value in the outskirts. Thus, the radial profiles are insensitive to the choice of the MAH model or the $c(M,z)$ relation. We note that the predicted evolution of the mass profile at the halo centre contrasts with the results found for individual halos in some cosmological $N$-body simulations \citep{Diemand2007}. More detailed studies are needed to achieve a firm conclusion on the evolution of the central mass structure of DM haloes. The main results of this paper do not depend on strongly on this behaviour, however, since the most significant orbital evolution of subhaloes occurs while subhaloes move in outskirts of the host halo, and the central mass is much smaller than the virial mass.

\bsp	
\label{lastpage}
\end{document}